\documentclass{revtex4}
\usepackage{graphicx}
\begin{document}
\bibliographystyle{apsrev}

\title{Gamma-Ray Summary Report}
\author{J. Buckley}
\email{buckley@wuphys.wustl.edu}
\affiliation{Washington University, 
St. Louis}
\author{T. Burnett}
\email{tburnett@u.washington.edu}
\affiliation{University of Washington}
\author{G. Sinnis}
\email{Gus@lanl.gov}
\affiliation{Los Alamos National Laboratory}
\author{P. Coppi}
\email{coppi@astro.yale.edu}
\affiliation{Yale University}
\author{P. Gondolo}
\email{pxg26@po.cwru.edu}
\affiliation{Case Western Reserve University}
\author{J. Kapusta}
\email{kapusta@physics.umn.edu}
\affiliation{University of Minnesota}
\author{J. McEnery}
\email{mcenery@titus.physics.wisc.edu}
\affiliation{University of Wisconsin}
\author{J. Norris}
\email{jnorris@lheapop.gsfc.nasa.gov}
\affiliation{NASA/Goddard Space Flight Center}
\author{P. Ullio}
\email{ullio@he.sissa.it}
\affiliation{SISSA}
\author{D.A. Williams}
\affiliation{University of California Santa Cruz}
\email{daw@scipp.ucsc.edu}

\date{\today}

\begin{abstract}
This paper reviews the field of gamma-ray astronomy and describes
future experiments and prospects for advances in fundamental physics
and high-energy astrophysics through gamma-ray measurements.
We concentrate on recent progress in the understanding of active galaxies, 
and the use of these sources as probes of intergalactic space. 
We also describe prospects for future experiments in a number 
of areas of fundamental physics, including: searches for an annihilation 
line from neutralino dark matter, understanding the energetics 
of supermassive black holes, using AGNs as cosmological probes 
of the primordial radiation fields, constraints on quantum gravity, 
detection of a new spectral component from GRBs, and the prospects 
for detecting primordial black holes.  

\end{abstract}

\maketitle

\section{Introduction}  

With new experiments such as GLAST and VERITAS on the horizon, we
are entering an exciting period for gamma-ray astronomy.  
The gamma-ray waveband has provided a new spectral window on
the universe and has already resulted in dramatic progress in 
our understanding of high energy astrophysical phenomena.
At these energies the universe looks quite different then when viewed 
with more traditional astronomical techniques.  
The sources of high energy gamma rays are limited to the most 
extreme places in the universe: the remnants of exploding stars,
the nonthermal Nebulae surrounding pulsars, the ultra-relativistic jets 
emerging from supermassive black holes at the center of active galaxies, 
and the still mysterious gamma-ray bursters.

While understanding these objects is of intrinsic interest
(how does nature accelerate particles to such high energies?
how do particles and fields behave in the presence
of strong gravitational fields?), these objects can also be
used as probes of the radiation fields in the universe and possibly
of spacetime itself.  In this case, the astrophysics of the object
is a confounding factor that must be understood to produce a
quantitative measurement or a robust upper limit.  While some may 
view this as a limitation of such {\it indirect} 
astrophysical measurements, in most cases there are no earth-bound experiments
that can probe the fundamental laws of physics at the
energy scales available to gamma-ray instruments.

Gamma-ray astronomy has developed along two separate paths. 
From the ground, simple, inexpensive experiments were built in the
1950's to observe the Cherenkov light generated by extensive air showers
generated by photons with energies above several TeV.  Despite decades
of effort it was not until the late 1980's that a source of TeV photons
was observed.  There are now roughly 10 known sources of TeV gamma rays,
three galactic sources and at least three active galaxies.
From space, the COS-B satellite, launched in 1975, observed the first
sources of cosmic gamma rays at energies above 70~MeV.  
The launch of the Compton Gamma Ray Observatory (CGRO) in 1991,
with the {\it Energetic Gamma Ray Experiment Telescope} (EGRET)
instrument, brought the field to maturity.
Whereas COS-B discovered a handful of sources, 
EGRET observed over 65 active galaxies\cite{Hartman99}, 
seven pulsars, many gamma-ray bursts,
and over 60 sources that have no known counterparts at other wavelengths. 
 
The disparity in the development of the two techniques can be traced
to the extremely low fluxes of particles present above a TeV
($\sim 4\gamma$ football field$^{-1}$~hr$^{-1}$) and the cosmic-ray 
background. Above the earth's atmosphere, one can surround a gamma-ray 
detector with a veto counter that registers the passage of charged particles.
From the ground, one is forced to infer the nature of the primary particle 
by observing the secondary radiation generated as 
the extensive air shower develops.  It was not until such a technique 
was developed for air Cherenkov telescopes~\cite{weekes-img},
that sources of TeV photons were discovered.  
Despite these difficulties a new generation of ground-based instruments 
is under development that will have a sensitivity that will rival that 
of space-based instruments.  At the same time a space-based instrument,
GLAST, with a relatively large area ($\sim 1$m$^2$) and excellent energy 
and angular resolution is scheduled to be launched in 2005.  

In this paper we will give a brief survey of the gamma-ray universe 
and demonstrate some of the fundamental measurements (relevant 
to particle physicists) that can be made using distant objects 
that emit high-energy photons.  What will hopefully become clear from this 
exposition are some development paths for future instruments.  
The need to see to the far reaches of the universe, makes a compelling 
case for ground-based instruments with energy thresholds as low as 10~GeV.  
The need to detect and study the many transient phenomena in the universe 
makes a compelling case for the development of an instrument that can 
continually monitor the entire overhead sky at energies above $\sim100$~GeV
with sensitivities approaching that of the next generation of pointed 
instruments.

As with any new branch of astronomy, it is impossible to predict what
knowledge will ultimately be gained from studying the universe 
in a different waveband, but early results hint at a rich future.  
New and planned instruments with greatly increased sensitivity will 
allow us to look farther into the universe and deeper into the 
astrophysical objects that emit gamma rays. Gamma-ray astronomy 
can be used to study the most extreme environments that exist in the 
universe, and may also provide a number of unique laboratories 
for exploring the fundamental laws of physics at energies beyond 
the reach of earth-bound particle accelerators.

\section{Physics Goals of Gamma Ray Astronomy}

\subsection{Active Galactic Nuclei}

Active galactic nuclei (AGN) are believed to be supermassive black holes,
$10^{8}-10^{10}M_{\odot}$, accreting matter from the nucleus of a host galaxy.
The accretion of matter onto a black hole is a very efficient process,
capable of releasing $\sim 10$\% of the rest energy the infalling matter
($\sim 40$\% for a maximally rotating black hole). (For comparison
fusion burning in stars releases $\sim0.7$\% of the rest energy.)
Radio loud AGN emit jets of relativistic particles, presumably along
the rotation axis of the spinning black hole.  The COS-B instrument 
observed the first AGN in the gamma-ray regime ($E>100$~MeV), 3C273.  
But it was not until the launch of the CGRO and EGRET that many AGN 
could be studied in the gamma-ray regime.  
More recently, ground-based instruments have extended these observations 
into the TeV energy band.
The energy output of these objects in gamma rays is of order 
$10^{45}$~ergs s$^{-1}$, and many of these objects emit most of their 
energy into gamma rays. 
The relativistic motion has several effects: 
1)~the energy of the photons is blue-shifted for an observer at rest (us), 
2)~the timescale is Lorentz contracted (further increasing the apparent 
luminosity), and 
3)~the relativistic beaming suppresses photon interactions.  
Thus, one expects that AGN observed in the TeV regime should have 
their jets nearly aligned with our line-of-sight.
 
The types of AGN detected at high energies, which include flat
spectrum radio quasars (FSRQs) and BL Lacertae (BL Lac) objects,
are collectively referred to as {\it blazars}.  
The Whipple Observatory 10m atmospheric Cherenkov telescope demonstrated 
that the emission spectra of several blazars extend into TeV energies.  
Two of these detections (Markarian 421 and Markarian 501) have been 
confirmed by independent experiments (CAT and HEGRA), 
at significance levels of between $20\sigma$ in a half hour to $80\sigma$ 
for a season.  

\begin{figure}[t] 
\noindent
\hskip -0.3in
\includegraphics[height=0.1\textheight]{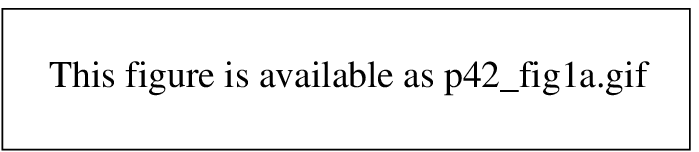}
\hskip -0.3in
\includegraphics[height=0.47\textheight]{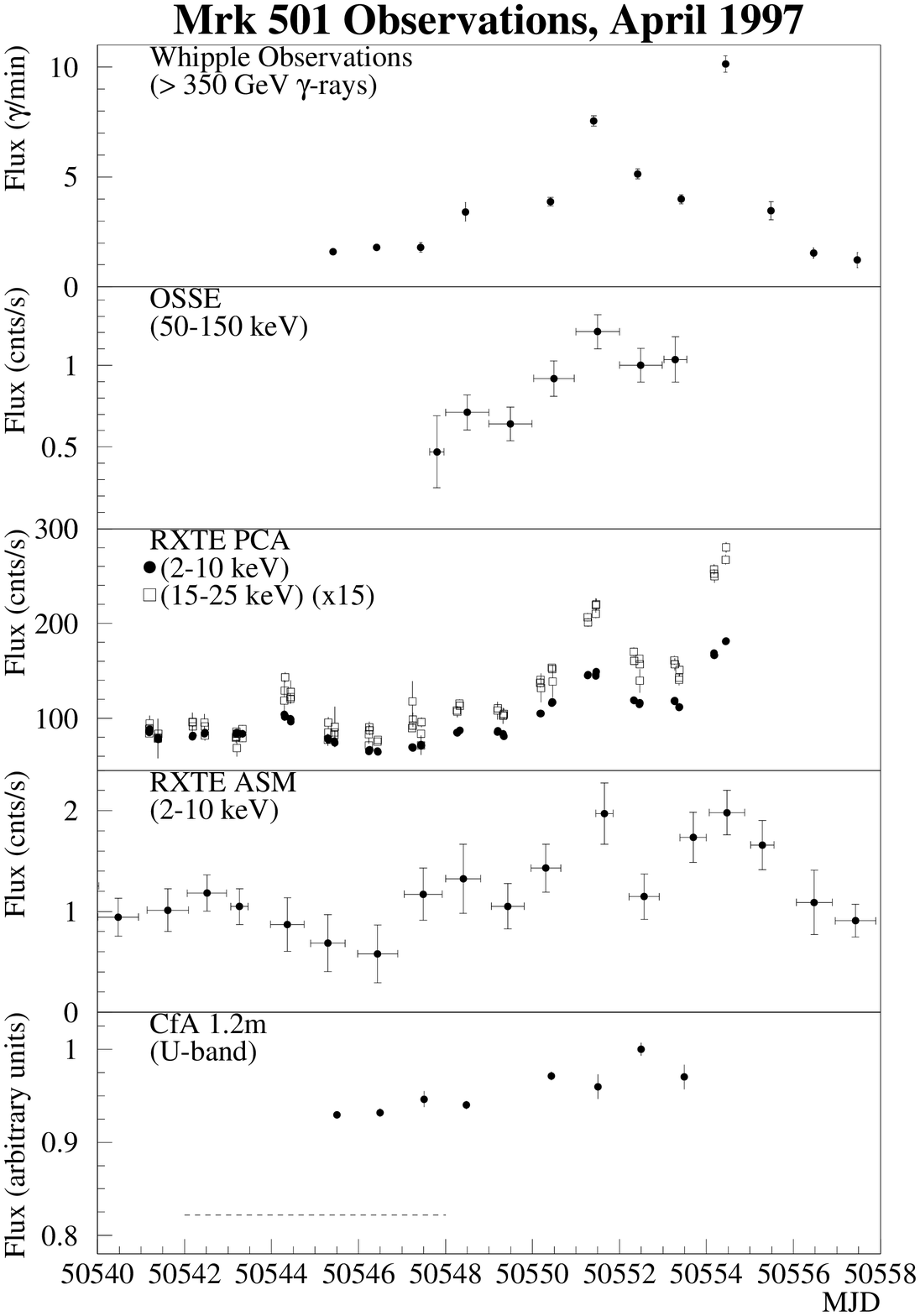}
\caption
{{\it Left:} SED of Mrk~501 from contemporaneous and archival observations. 
{\it Right:} Multi-wavelength observations of Mrk 501; 
(a)~$\gamma$-ray, (b)~hard X-ray, (c)~soft X-ray, 
(d)~U-band optical light curves during the period 1997 April 2--20 
(April 2 corresponds to MJD 50540).  
The dashed line in (d) indicates the optical flux in 1997 March.
(from \cite{Buckley01} and references therein.)
\label{fig:m5nufnu}}
\end{figure}

Blazar emission is dominated by highly variable,
non-thermal continuum emission from an unresolved nucleus.
The broadband emission and high degree of polarization suggest
synchrotron radiation extending from radio up to UV or even hard
X-ray energies.
The short variability timescales and high luminosities are thought
to result from highly relativistic outflows along jets pointed very
nearly along our line of sight.
The spectral energy distributions (SEDs) of these objects
have a double-peaked shape (see Figure~\ref{fig:m5nufnu}) with 
a synchrotron component that peaks in the UV or X-ray band, 
and a second component typically rising in the X-ray range 
and peaking at energies between $\sim 1$~MeV and 1~TeV~\cite{Montigny95}.  
The most natural explanation of the second peak is inverse-Compton
scattering of ambient or synchrotron photons~\cite{Sikora94}
although other possibilities such as proton-induced cascades
have not been ruled out~\cite{mannheim93}.  
These two models have somewhat complementary strengths and weaknesses.  
Since electrons are lighter than protons, they can be confined
in a smaller acceleration region but lose energy more quickly (by
synchrotron and IC emission), making it
difficult to accelerate electrons 
to extreme energies.  For hadronic models, 
very high energies can be attained given sufficient time, a large
acceleration region and high magnetic fields. However, the short 
variability timescales, implying short acceleration times and compact
regions are difficult to explain.  In addition, the electron models 
make natural predictions on the correlation between X-ray 
and gamma ray luminosities.  While it has been claimed that proton 
models can be constructed that explain these correlations,
detailed calculations have not appeared in the literature.

Whipple observations of the vast majority of EGRET blazars have
yielded only upper limits~\cite{Kerrick95b,Quinn95,Petry97}; 
Mrk~421 ($z =0.031$)~\cite{Punch92} being the exception.
Subsequent searches for emission from X-ray bright BL Lac objects
has led to the detection of Mrk~501 ($z=0.034$)~\cite{Quinn96}, 
and four other as yet unconfirmed sources 
[1ES~2344+514 ($z=0.044$~\cite{Catanese98},
1ES~2155-304 ($z=0.117$)~\cite{chadwick99},
1ES1959+650 ($z=0.048$)~\cite{Nishiyama99} and 
1H1426+428 ($z=0.13$)~\cite{Horan01}]. 
The SEDs observed for these sources show higher energy synchrotron 
and $\gamma$-ray peaks, and comparable power output at the synchrotron 
and $\gamma$-ray peak.

These observations are well described by the classification scheme of
Padovani and Giommi~\cite{padovani95}.  
The AGN detected by EGRET are all radio-loud, flat-spectrum radio 
sources and lie at redshifts between 0.03 and 2.28.  
They are characterized by two component spectra with peak power 
in the infrared to optical waveband and in the 10 MeV to GeV range.  
For many of the GeV blazars, the total power output of these sources 
peaks in the gamma-ray waveband.

The objects detected at VHE, appear to form a new class
distinct from the EGRET sources.  
All are classified as
{\it high-energy peaked~\cite{padovani95} BL Lacs (HBLs)} defined 
as sources with their synchrotron emission peaked in the UV/X-ray band 
and gamma-ray emission peaking in the $\sim$100 GeV regime
(see, e.g., Fig.~\ref{fig:m5nufnu}).
The correspondence of the position of the peak of the synchrotron
and $\gamma$-ray energy is naturally explained in models where the
same population of electrons produces both spectral components.  
Proton induced cascade models~\cite{mannheim93} might also reproduce 
the spectra, but have no natural correlation in the cutoff energy 
of the two components, or the observed correlated variability. 

Another difference in the VHE detections is that only the nearest
sources with redshifts $z \lesssim 0.1$ have been detected.
The sensitivity of EGRET for a one-year exposure is comparable 
to that of Whipple for a 50 hour exposure for a source with spectral 
index of 2.2.  The failure of ACTs to detect any but the nearest AGNs 
therefore requires a cut-off in the $\gamma$-ray spectra of the EGRET 
sources between 10~GeV and a few hundred GeV.  This cutoff could be 
intrinsic to the electron acceleration mechanism,
due to absorption off of ambient photons from the accreting nuclear
region~\cite{Dermer92b}, or caused by absorption via pair production 
with the diffuse extragalactic background radiation~\cite{Gould67,Stecker93}.
While the latter mechanism establishes an energy-dependent 
gamma-ray horizon it can also be used to measure the
radiation fields that fill intergalactic space.  

In the framework of Fossati et al.,~\cite{fossati98}
the low energy peaked EGRET BL Lacs (LBLs) correspond to AGNs with a
more luminous nuclear emission component than HBLs.  
The relatively high ambient photon density in the LBLs 
is up-scattered by relativistic electrons to $\gamma$-ray energies.  
With high enough ambient photon densities, the resulting 
inverse-Compton emission can exceed that resulting from the
up-scattering of synchrotron photons. 
This accounts for the observation of relatively high
levels of gamma-ray emission, dominating the power output over 
the entire spectrum.

The higher luminosity could also shut down the acceleration process
at lower energies.  
For lack of another viable hypothesis, consider the common hypothesis that
the energetic particles in AGNs come from electrons or protons 
accelerated by relativistic shocks traveling down the AGN jets.
In the model of diffusive shock acceleration 
(essentially the first order Fermi process), particles are 
accelerated as they are scattered from magnetic irregularities 
on either side of a shock.  For strong, non-relativistic shocks,
a constant escape probability with each shock crossing results 
in an $\sim E^{-2}$ spectrum, close to that observed.  
More realistic models including nonlinear effects
lead to slightly steeper spectra; if the shock velocity is relativistic 
the spectral index may range from 
1.7 to 2.4.
In any event, an electron spectrum $\sim E^{-\gamma}$ will give rise 
to synchrotron radiation with a spectral index $\alpha =(\gamma -1)/2$, 
in good agreement with observations.

The maximum energy attainable is given by equating the rate 
of energy loss from synchrotron emission or inverse-Compton
emission to the acceleration rate as given by the shock parameters.  
In the low-energy peaked objects, it is thought that high ambient
photon densities result in inverse-Compton losses that dominate 
over synchrotron losses and limit the maximum electron energy 
achieved by shock acceleration. 
Thus one also obtains a natural explanation for the lower energies 
of the peak synchrotron and IC power in these objects.  
In HBLs, the ambient photon fields are presumably weaker and 
self-Compton emission dominates over Comptonization of external 
photons (EC).  Electrons can reach higher energies by shock acceleration, 
and the peaks in the SED move to higher energies and have more nearly 
equal peak power.
This model is consistent with the data and serves as a useful paradigm 
for searching for new VHE sources.

The SEDs shown in Fig.~\ref{fig:m5nufnu},
combine the results of a number of different
measurements of the X-ray and VHE spectra of
Mrk~501, and compare them with simple synchrotron self-Compton (SSC) 
models (see Buckley ~\cite{Buckley01} and references therein). 
The agreement between the spectral measurements 
and the model is exceptionally good for Mrk~501.

\vspace*{-0.1in}
\subsubsection{Multiwavelength Observations: Variability}
\label{sec:variability}

\begin{figure}[t]
\begin{center}
\includegraphics[height=0.1\textheight]{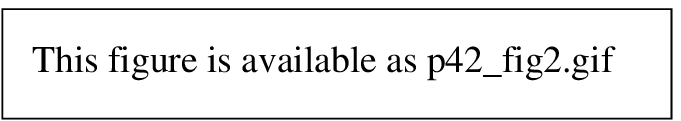}
\end{center}
\caption{
Simultaneous X-ray/ $\gamma$-ray flare observed on March 19, 2001.
The 2--10~keV X-ray light curve was obtained with the PCA detector 
on RXTE~\protect\cite{Jordan01,Fossati01};
data points are binned in roughly 4 minute intervals.
\label{fig:newmulti}}
\end{figure}

Data taken on Mrk 421 over the years 1995~\cite{Buckley96} 
to 2001~\cite{Jordan01} show that the $\gamma$-ray emission is
characterized by a succession of approximately hour-long flares with relatively
symmetric profiles (see Figure~\ref{fig:newmulti}).

While most of the multiwavelength observations of Mrk~421 show
evidence for correlated X-ray and gamma ray variability, the nature
of the correlation is unclear and the data have traditionally undersampled
the variability.  However, a multi-wavelength campaign conducted 
on Mrk~501 in 1997 revealed a strong correlation between TeV 
$\gamma$-rays and soft X-rays (the 50--500~keV band detected 
by OSSE) (Fig.~\ref{fig:m5nufnu}). 

Recent multiwavelength observations of Mrk~421 made during the period 
March 18, 2001 to April 1, 2001 with the Whipple gamma-ray telescope, 
and the Proportional Counter Array (PCA) detector on the Rossi
X-ray Timing Explorer (RXTE)  
better sample the rapid variability of Mrk~421.  Key to the
success of this campaign is the 
nearly continuous $>330$~ks exposure with RXTE~\cite{Fossati01}.
Numerous ground-based atmospheric Cherenkov and optical observations 
were scheduled during this period to improve the temporal coverage 
in the optical and VHE bands.
Frequent correlated hour-scale X-ray and $\gamma$-ray flares were observed.
Fig.~\ref{fig:newmulti} shows a subset of these data showing the close 
correlation of the well-sampled TeV and X-ray (2--10~keV)
lightcurves on March 19, 2001~\cite{Jordan01}.

Leptonic models provide a natural explanation
of the correlated X-ray and gamma-ray flares, and can
reproduce the shape of the flare spectrum.
The simplest model for blazar emission is the one-zone synchrotron 
self-Compton (SSC) model where energetic electrons in a compact emission 
region up-scatter their own synchrotron radiation.  As shown in
Fig.~\ref{fig:m5nufnu}, such a model results in surprisingly good fits 
to the Mrk~501 SED.   In the SSC model, the intensity of the synchrotron 
radiation is proportional to the magnetic energy density and the number 
density of electrons $I_{\rm synch} \propto n_e$.
Since these same electrons up-scatter this radiation, the IC emission 
scales as $I_{\rm IC} \propto n_e^2$.  Thus we expect 
$I_{\rm IC} \propto I_{\rm synch}^2$.

\begin{figure}
\begin{center}
\includegraphics[height=.34\textheight]{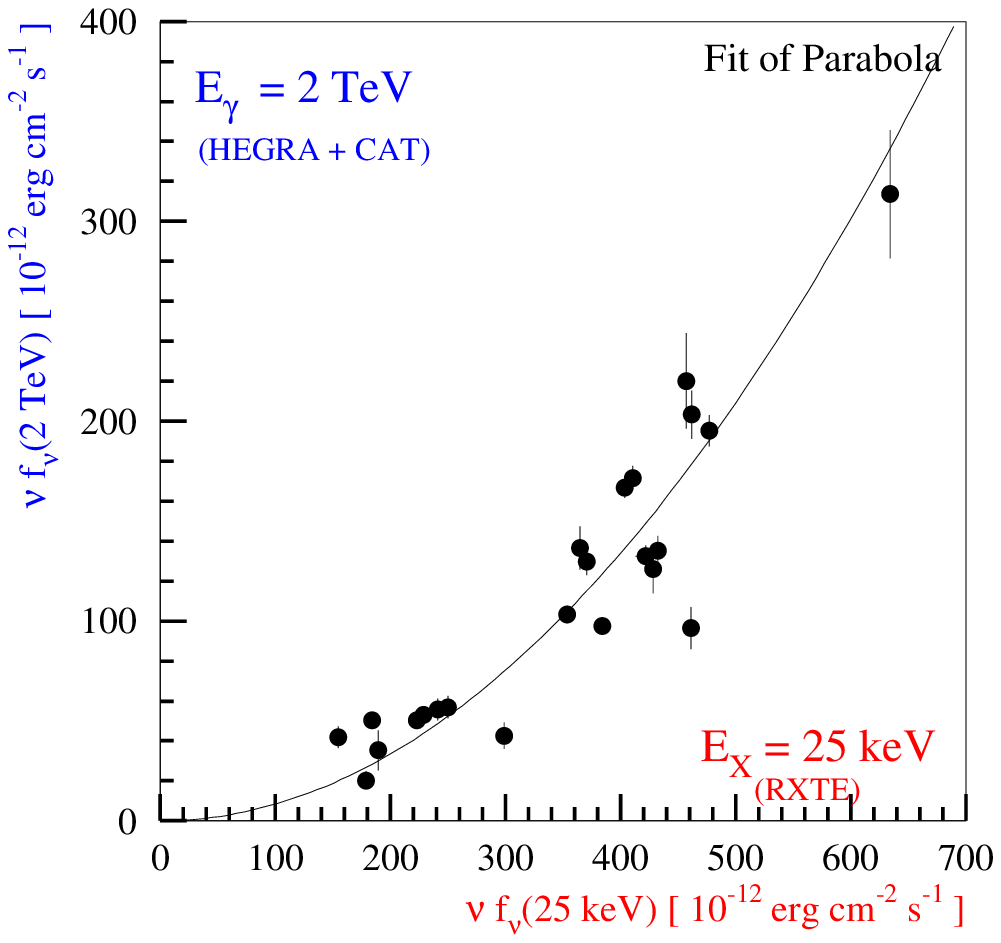}
\end{center}
\caption{Plot of TeV $\gamma$-ray flux versus X-ray flux measured
with the HEGRA experiment during an intense flare of Mrk~501
(courtesy Henric Krawczynski).}
\label{fig:tvx}
\end{figure}

Krawczynski et al.,~\cite{Krawczynski00} examined the correlation 
of TeV $\gamma$-ray and X-ray intensity for several strong flares 
of Mrk~501 in 1997.  The results, plotted in Figure~\ref{fig:tvx},
show evidence for such a quadratic dependence. 
(However the possibility of a baseline level of the X-ray emission
can not be excluded.)  While the interpretation of
these observations is not unambiguous, this analysis
is an important example
of what can be learned with continued multiwavelength studies of
AGNs. 

How do these observations constrain the alternative hypothesis that
proton induced cascades (PIC), not electrons, are responsible for the 
gamma-ray emission?
In the hadronic models of Mannheim and collaborators,
the gamma-ray emission typically comes from synchrotron 
emission from extremely energetic, secondary electrons produced 
in hadronic cascades.  Since a viable hadronic target for 
$pp\rightarrow pp\pi$ appears to be lacking (except perhaps in 
the broad line clouds), the assumption is made that the cascade begins 
with ultrarelativistic particles interacting with ambient photons 
to produce pions.   This implies proton energies in excess of 
$10^{\sim 16}$~eV.  The neutral pions presumably give rise to 
gamma rays and electromagnetic cascades, while the charged pions 
could give a neutrino signal.  These models have attracted much
interest since, in the most optimistic cases, these models may produce
an observable neutrino signal and may provide a mechanism for producing
the ultra-high energy cosmic rays.  If the sources are 
{\it optically thick} to the emerging protons 
(i.e., they absorb some fraction of the cosmic rays, 
but not the neutrinos) then it may be possible to produce 
a relatively large neutrino signal without overproducing the local
cosmic ray flux
\cite{waxman-bahcall-01}.  While these models have a number of
attractive features, there is some debate about whether they can provide
a self-consistent description for the observations.

To overcome the threshold condition for pion production, 
protons must have energies in excess of $10^{16}$ to $10^{18}$~eV where
abundant infrared photons can provide the target.
Since the cross section for photo-pion production is relatively low, 
very high ambient photon densities are required to initiate the cascades.  
In this case, pair creation ($\gamma\gamma\rightarrow e^+ e^-$), 
which has a much higher cross-section, must be important. 
The proton cascade models may well have a significant problems explaining
the emission from objects like Mkn 421/501 for this reason.
In the PIC models \cite{mannheim93} the proton-photon
interaction occur with radio-IR photons
in the jet.  While a detailed analysis has not been published,
Aharonian and others have pointed out that the
required photon densities also imply large pair
production optical depths, and may mean that the PIC models are not
self-consistent.  Models where the primary protons produce synchrotron
radiation (and subsequent pair-cascades) may avoid this problem,
but require even larger magnetic fields \cite{Aharonian00}.

One advantage of the photon-pair cascade is that it produces a
rather characteristic spectrum that does not depend sensitively
on the model parameters.  The detailed shape of this spectrum
does not match some observations.  Typically the
spectra are too soft and overproduce
X-rays, giving a spectrum that does not reproduce the strongly double-peaked
spectrum observed.  For the typical magnetic field values, the synchrotron
spectrum is often too soft and lacks the spectral breaks that are 
observed. 

For these hadronic models to account for the double-peaked
spectrum, the radio to X-ray emission is most likely
produced by primary shock-accelerated electrons, while the gamma-ray
emission is produced by energetic secondary electrons from the cascade.
There is no natural explanation for the correlated variability in the
two spectral bands, or in the correlation in the X-ray and gamma-ray
cutoff energy.

To reach these energies on a sufficiently short timescale,
the gyroradius must be limited to a compact region in the jet, the 
inverse-Compton emission must be suppressed, and magnetic fields of up
to 40~Gauss are required.  The spectral variability seen in the X-ray
waveband is consistent with much longer synchrotron cooling times
than predicted by the hadronic models, and is quite consistent with
magnetic fields of a 10 to 100 mGauss.  This is the
same value of the magnetic field derived by a completely independent
method within the framework of the synchrotron inverse-Compton model.

The criticisms leveled at the electron models are that the magnetic
fields are too small compared with the value required for magnetic
collimation of the jets, and that the required electron energies 
are too large to be explained by shock acceleration.  
Moreover, electron injection into shocks is poorly understood 
since the electron gyroradius is small compared to the proton 
gyroradius and presumably to the width of the broadened shock front.  
However we know that electrons are accelerated to 100 TeV
energies in supernovae shocks,
regardless of the theoretical difficulties in accounting for this
observation.  As will be shown below, if one accepts relatively large
Doppler factors, a self-consistent explanation for the VHE gamma-ray
emission can be derived from leptonic models.

\begin{figure}[t!]
\begin{center}
\includegraphics[width=0.7\textwidth]{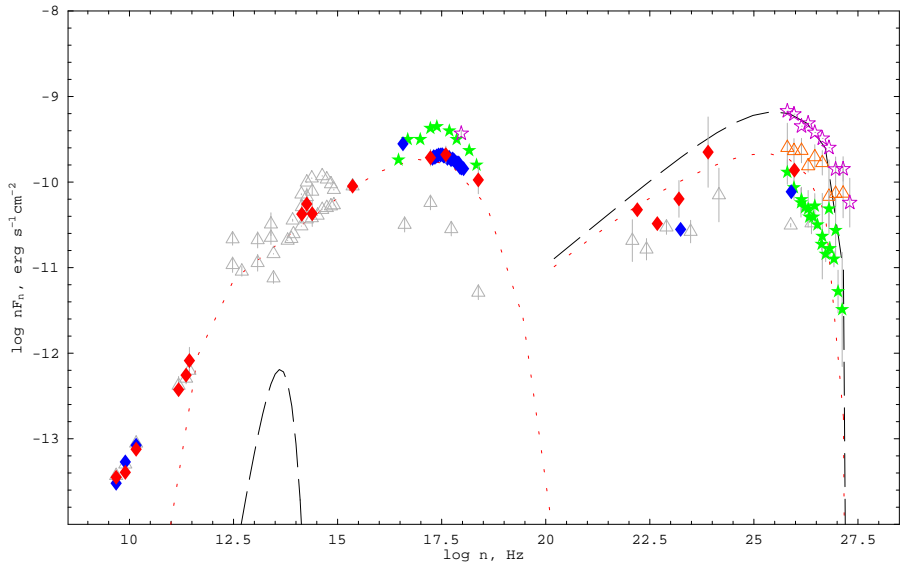}
\end{center}
\caption{Model fit to Mrk~421 SED with both an SSC and external 
Compton component\cite{Buckley01}} 
\label{fig:m4sedfit}
\end{figure}

In the framework of either the EC or SSC models the $\gamma$-ray 
and X-ray data can be used to constrain the Doppler factor 
$\delta$ (this is thought to be close to the bulk Lorentz factor 
of the jet for blazars) and magnetic field $B$ in the emission 
regions of Mrk~421 and Mrk~501.
The maximum $\gamma$-ray (IC) energy $E_{\rm C,max}$ provides 
a lower limit on the maximum electron energy (with Lorentz factor 
$\gamma_{\rm e,max}$) given by 
$\delta \gamma_{\rm e,max} > E_{\rm C,max}/ m_e c^2$;
combining this with the measured cut-off energy of the
synchrotron emission $E_{\rm syn,max}$
one obtains an upper limit on the magnetic field 
$B\lesssim 2\!\times\! 10^{-2} E_{\rm syn,max} \delta E_{\rm C,max}^{-2}$ 
(where $E_{\rm C,max}$ is in TeV).  A lower limit on the magnetic field 
follows from the requirement that the electron cooling time,
$t_{\rm e,cool} \approx 2\!\times\! 10^8\delta^{-1}\gamma_{\rm e}^{-1}
B^{-2} {\rm s}$, must be less than the observed flare decay timescale.
These limits depend on the Doppler factor of the jet and in some cases 
cannot be satisfied unless $\delta$ is significantly greater than 
unity ~\cite{Catanese97a,Buckley97}.
Typically, these arguments lead to predictions of $\sim 100$~mGauss 
fields and Doppler factors $\delta > 10$ to 40 for Mrk~421. 
Similar values for Mrk~501 but typically with a reduced lower limit 
on the Doppler factor.  Model fits (that ignore the fact that the
multiwavelength data are not truly time-resolved) give similar
values for the Doppler factor and magnetic field strength.  For example,
a simple one-zone model fit for Mrk~421, shown in Fig.~\ref{fig:m4sedfit},
only gives good fits for a Doppler factor approaching
a value of $\delta \approx 100$ (as shown) \cite{Buckley01}.

Doppler factors this large may present other problems.  Radio observations
of jets show radio components moving with velocities that imply bulk
Lorentz factors $\Gamma \lesssim 10$ further out in the jet.  If the
jet is decelerated by the inverse-Compton scattering, 
most of the energy would be used up before such extended radio lobes 
could form in apparent contradiction to observations.

Given the good progress to date, it appears that it will 
be possible to determine the dominant radiation processes in
AGNs.  After this first issue is resolved, further multiwavelength
observations can address the more fundamental questions about the energetics
of the central supermassive black hole, and the processes behind
the formation of the relativistic jets.  The very short variability timescales
already observed with the Whipple instrument (15 minute doubling times for
Markarian 421) hint that the gamma-ray observations may be probing very
close to the central engine, beyond the reach of the highest resolution
optical and radio telescopes.

\subsection{Gamma-Ray Bursts}

\begin{figure}[t]
\label{fig:grbdistance}
\begin{center}
\includegraphics[height=0.1\textheight]{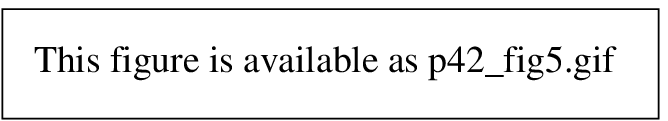} 
\end{center}
\caption{The magnitude redshift distribution of gamma-ray bursts.
Also shown on the plot
is the magnitude vs. redshift relation for the observed type Ia supernovae.}
\end{figure}

Gamma-ray bursts (GRBs) were discovered by the Vela satellites in the
late 1960's~\cite{klebs}.  GRBs are bright flashes of hard X-rays and
low energy gamma rays coming from random directions in the sky at 
random times.  Until the launch of the CGRO in 1992 it was generally 
believed that GRBs were galactic phenomena associated with neutron stars.  
The BATSE instrument on-board the CGRO detected over 2000 GRBs and the 
observed spatial distribution was isotropic, with no evidence of 
an excess from the galactic plane.  Thus GRBs were either cosmological 
or populated an extended galactic halo.  In 1997 the BeppoSax satellite 
was launched.  With a suite of hard X-ray detectors, this instrument 
has the ability to localize GRBs to within $\sim 1$ minute of 
arc~\cite{costa} (BATSE could localize GRBs to within $\sim 5$ degrees). 
The increased angular resolution allowed conventional ground-based
telescopes to search the error box without significant source confusion.
The observation of emission and absorption lines
from the host galaxies led to measurements of redshifts;
some thirty years after their discovery the cosmological
nature of gamma-ray bursts was determined.  In Figure~\ref{fig:grbdistance}
we show the redshift distribution of those gamma-ray bursts where the
redshift has been determined.  The enormous energy output from GRBs, 
and transparency of the universe below 100~MeV makes GRBs visible across 
the universe.  Thus gamma-ray bursts have the potential to probe 
the universe at very early times and to study the propagation of 
high-energy photons over cosmological distances.

To use GRBs as cosmological probes it is necessary to understand
their underlying mechanism.  While GRBs may never be standard candles
on par with the now famous Type-IA supernovae, there has been great
progress made in the last five years in understanding GRBs.  
While we still do not know what the underlying energy source is, 
we are beginning to understand the environment that creates 
the observed high-energy photons.

The large distances to GRBs implies that the energy released is
$\sim10^{50-54}$ ergs, depending on the amount of beaming at the source.  
While the origin of the initial explosion is unknown, the subsequent 
emission is well described by the relativistic fireball model.  
In this model 
shells of material expand relativistically into the interstellar medium.  
The complex gamma-ray light-curves of the prompt radiation arises 
from shocks formed as faster and slower shells 
of material interact.   A termination shock is also formed as 
the expanding shells of material interact with the material surrounding 
the GRB progenitor.
In this model the observed afterglows (x-ray, optical, and radio) 
arise from the synchrotron radiation of shock accelerated electrons.

The afterglow emission can be used to determine the geometry of the source.  
Since the shell is expanding relativistically, the radiation  
(emitted isotropically in the bulk frame) is beamed into a cone with
with opening angle $\Gamma^{-1}$ (the bulk Lorentz factor of the 
material in the shell).  Thus at early times, only a small portion 
of the emitting surface is visible and one cannot distinguish between 
isotropic and beamed (jet-like) emission.
However, as the shell expands it sweeps up material and $\Gamma$ decreases.  
If the emission is not isotropic the beaming angle ($\Gamma^{-1}$) 
will eventually become larger than the opening angle of the jet.  
At this point one should observe a break in the light curve 
(luminosity versus time) of the afterglow.  
This distinctive feature has been observed in 15~GRBs.  
By measuring the temporal breaks in GRBs of known redshift 
Frail et al.,~\cite{frail} have measured the jet opening angles of
15 gamma-ray bursts (with some assumptions about the emission region:
the jet is uniform across its face, the electron distribution in the
shock is a power law, the afterglow radiation is due to synchrotron
emission and inverse Compton scattering).  
If one integrates the observed luminosity over the inferred jet opening 
angle one can determine the intrinsic luminosity of each GRB.
Surprisingly, Frail et al., conclude that the intrinsic luminosities
of the observed gamma-ray bursts are peaked around 
$5\times10^{50}$~ergs with a spread of roughly a factor of six.
Thus the observed variation in luminosity (a factor of $\sim 500$) 
may be mainly due to the variation in the jet opening angle.
Note that this conclusion applies only to the ``long" GRBs, 
as these are the only GRBs for which optical counterparts
have been observed.  

\begin{figure}[t]
\label{fig:grbhr}
\begin{center}
\includegraphics[height=.1 \textheight]{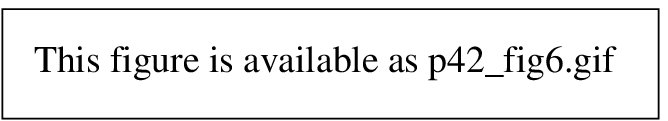}
\end{center}
\caption{Observed GRB luminosity, assuming isotropic emission vs. observed
lag between high and low energy photons in the BATSE instrument.
Channel 1 includes photons with energies between 25 and 50~keV 
and channel 3 includes photons between 100--300~keV.}
\end{figure}

With a similar goal, to reduce the wide divergence in the observational
properties of GRBs, Norris~\cite{norris99} has found a correlation 
between energy dependent time lags and the observed burst luminosity.  
Three things occur as one moves from high energy photons 
to low energy photons.  
The pulse profiles widen and become asymmetric, and the centroid
of the pulse shifts to later times.  The time lag is defined 
as the shift in the centroid of the pulse profile in the different 
energy channels of the BATSE instrument.  In Figure~\ref{fig:grbhr} 
we show the observed luminosity (assuming isotropic emission) versus 
the time lag observed between two energy channels on the BATSE experiment. 
(Channel 1 corresponds to photons with energies 
between 25--50~keV and channel 3 to 100~300~keV photons.) 
The line is the function, 
$L_{53}=1.1\times(\tau_{lag}/0.01s)^{-1.15}$, where $L_{53}$ is
the luminosity in units of $10^{53}$~ergs.  It may be that the time lag
is dependent upon the jet opening angle for reasons that are not yet 
understood and this observed correlation is simply an way 
of paramterizing the relationship observed by Frail et al.

As discussed above, gamma-ray observations of AGNs revealed a new 
spectral component due to inverse-Compton emission, 
distinct from the synchrotron emission observed in the radio to X-ray 
wavebands.  This observation resulted in an independent constraint 
on the electron energy that allowed a determination of the magnetic fields, 
electron densities, and bulk Lorentz factors in the sources.
While AGNs are quite different for GRBs, the non-thermal radiation mechanisms
may be quite similar, and we might expect similar progress to follow from
high energy gamma-ray measurements.

At higher energies less is known about GRBs.   The EGRET instrument 
covered the energy range from 100 MeV to a few tens of GeV.  EGRET 
detected several GRBs at high energy (HE $E>100$~MeV).
From EGRET observations we know that the energy spectrum is a power 
law that extends to high energies $\sim$GeV.  
However, dead-time in the EGRET spark chambers compromised the ability 
of EGRET to measure light curves and spectra.
  
The high energy spectrum is of inherent interest in understanding 
the environment of the GRB.
Observations above MeV energies provide important constraints on the
Doppler factor. Without a large Doppler factor, photon-photon pair
production interactions would cut off the photon spectrum. 
The highest energy photon observed from a GRB was seen by EGRET.  
They detected an 18~GeV photon from a GRB~\cite{hurley}
that arrived roughly 90~minutes after the lower energy photons 
(25--150~keV observed by Ulysses).  There has also been tantalizing 
evidence of even higher energy photons from gamma-ray bursts.  
The Milagrito detector, a ground-based gamma-ray telescope, 
observed an excess of events (18 observed on a background of 3.46 events)
in spatial and temporal coincidence with GRB~970417a.  
However, due to the poor localization of the BATSE detection and 
the large number of bursts examined (54), the detection is marginal 
($\sim 3 \sigma$).  If the observed excess is not due to a statistical 
fluctuation the implied photon energies from the burst are $\gtrsim 200$~GeV. 

\begin{figure}[t]
\label{fig:grb_sense}
\begin{center}
\includegraphics[width=9.5cm]{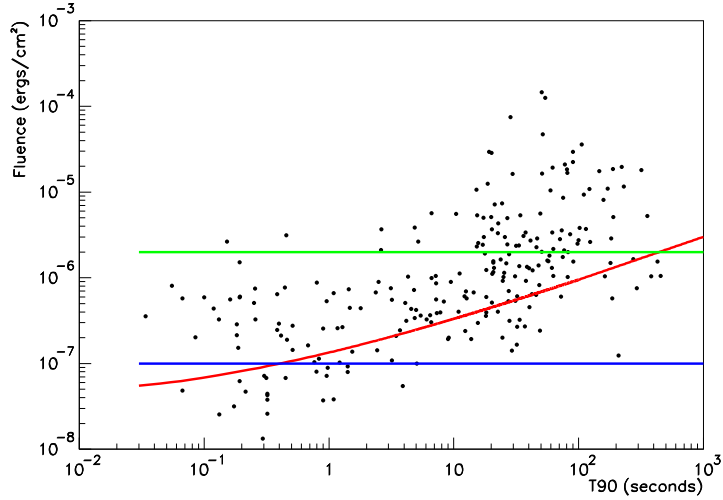}
\end{center}
\caption{Fluence sensitivity of EGRET~(green line) Milagro~(red) 
and GLAST~(blue) to GRBs as a function of burst duration.  
The data points show the distribution of fluence and duration 
seen by the BATSE instrument.}
\end{figure}

There are new instruments both running and planned that should greatly 
increase the sample of high-energy photons seen from gamma-ray bursts.
Figure~\ref{fig:grb_sense} shows the sensitivities of EGRET, Milagro,
and GLAST.  The sensitivity plotted for Milagro assumes that the source 
spectrum continues to TeV energies and is unaffected by intergalactic 
absorption (GRBs with a redshift beyond $\sim$0.3--0.5 should not be 
visible to Milagro).  

\begin{figure}[t]
\label{fig:veritasgrb}
\begin{center}
\includegraphics[width=10cm]{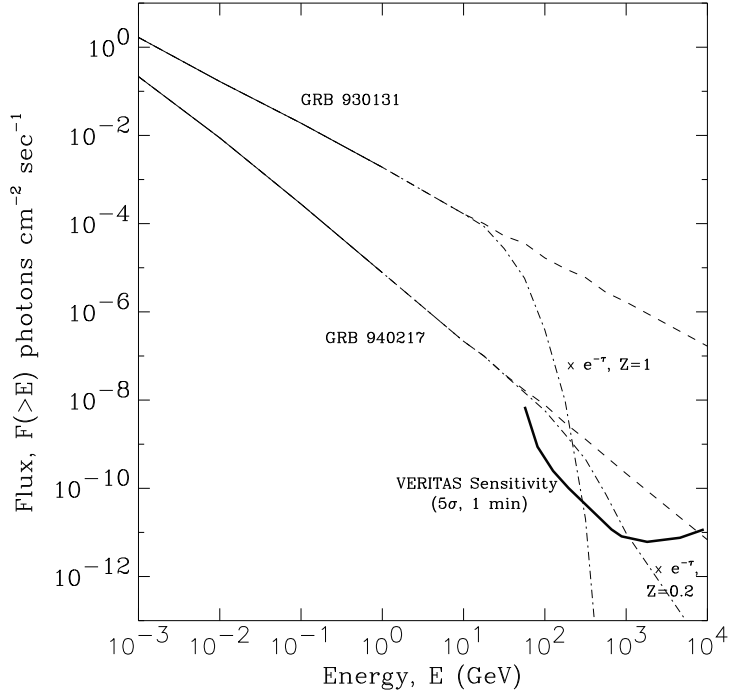}
\end{center}
\caption{Sensitivity of VERITAS to two hypothetical bursts extrapolated
from EGRET detections.}
\end{figure}

VERITAS will also have sensitivity to GRBs if a TeV afterglow persists for a 
few minutes and the burst has been localized to within
$\sim 1$~degree, so the telescopes can slew to the source.
In Figure~\ref{fig:veritasgrb} we show power-law extrapolations of two 
EGRET detected GRBs using the measured spectral indicies.  Since no redshifts
were available for these sources, for illustration we calculate the effect
of infrared absorption on one source assuming a moderately large redshift
($Z=1$) and for the other source assuming a redshift closer to the minimum
value determined for a GRB. 

All of these instruments will have vastly increased sensitivity to the 
high energy spectrum of gamma-ray bursts, and we should be able to
better understand the environments (bulk Lorentz factor, magnetic fields,
and particle densities) around the GRB.

Wide field instruments like GLAST and Milagro have natural advantages 
for detecting GRBs.
While Cherenkov detectors suffer from a small field of view,
and small limited duty cycle (about 10\% corresponding to dark moonless night) 
compared with MILAGRO and GLAST, their relatively large effective 
areas above $\sim$50~GeV should prove important for a small fraction of GRBs.
The sensitivity of air Cherenkov telescopes is good enough that
even for a source at a redshift of $z\approx 1$, with several e-foldings
of attenuation, a strong detection could be possible.
If the VHE emission persists long enough 
to be detected, the huge effective area of ground-based telescopes such
as VERITAS would provide as many as $10^4$ times the number of detected photons
above 100 GeV as space-based experiments, greatly enhancing the spectral
reach of GLAST.

\subsection{Search for Dark Matter}

One of the most exciting prospects for future gamma-ray experiments
is the possibility of detecting the unambiguous signature of cold dark 
matter in the galaxy through the annihilation of 
these particles to a gamma-ray line.
However, as most will have surmised from this Snowmass workshop,
the standard Cosmological model has been in a dramatic state of upheaval in 
recent years.  For this reason it is worth first revisiting the case 
for cold dark matter, then looking in detail at the prospects for a
gamma-ray detection.

Over the last couple of years our knowledge 
of the inventory of matter and energy in the universe has improved 
dramatically.  Astrophysical measurements from disparate experiments 
are now converging and a standard cosmological model is emerging.
The most significant new data come from recent 
measurements of the cosmic microwave background radiation (CMBR) 
and measurements of the Hubble flow using distant supernovae.

Temperature fluctuations in the CMBR trace the matter distribution
at the time that electrons and protons combined to form atoms, and
matter and radiation decoupled.  Measurements of the power spectrum
of the fluctuations tell us that, the density of the universe is close
to the critical density ($\Omega_{\rm tot} \approx 1$) and that matter,
both luminous and dark, makes up at most 30\% of the total.  

Type IA supernovae have been detected out to a redshift of 1.7.
Despite the intuitive idea that gravity should be causing the universal 
expansion to slow down, the luminosity redshift relationship shows 
evidence for a slower expansion in the past than in the present.  
This {\it acceleration} can only be explained by a form of dark energy 
with a strange equation of state that results in a net repulsive force,
where the total dark energy 
density is given by $\Omega_\Lambda \approx 0.7$. 
How does this discovery affect our expectations about the matter density
of the universe?

It is now well established from the observation of galactic rotation 
curves that our galaxy is comprised of a dark halo of density 
$\Omega_{\rm halo} > 0.1$ which extends well beyond the 
distribution of visible matter ($\Omega$ is defined as the ratio 
of the density $\rho$ to the critical density $\rho_{\rm crit}$ 
for which the universe is flat).
Large scale motions of galaxies on the scale of clusters and
superclusters indicate a still higher value of $\Omega\gtrsim{0.2}-0.3$
(in good agreement with a $\Omega_{\rm total} - \Omega_{\Lambda}$).
However, an inventory of the stars shows that only about 10\% of this 
total is in the form of luminous matter.  Much effort has gone into
determining the nature of this unseen, or dark, matter.  

Until recently,
the best handle on the amount of ordinary (baryonic) matter came not
from starlight, but from
measurements of the primordial abundances of deuterium, 
$^3$He, $^4$He and $^7$Li.  Detailed calculations of nucleosynthesis 
in the very early universe require that the total amount of matter 
in the form of baryons is 
$0.008 \lesssim \Omega_{\rm baryon} h^2 \lesssim 0.024$, or conservatively
$\Omega_{\rm baryon} \lesssim 0.1$.  More recently,
the position and intensity of the second acoustic peak in the CMBR
has been used to independently determine the total baryon density.  
This peak in the CMBR power spectrum is determined by the balance 
between the restoring force of photon pressure on baryons and 
the gravitational force, giving a new measure of the baryonic density
of the universe.
The two measurements agree amazingly well, providing further support
for a standard cosmology, but revealing a large component of the matter
density of the universe that is non-baryonic and is distinct
from the {\it dark energy}.

Moreover, if most of the matter in the 
universe is non-baryonic it is reasonable to assume that much of the
matter comprising our own galactic halo is non-baryonic as well.  
The origin of structure relatively early in the history of the 
universe strongly favors cold dark matter (matter that was 
relativistic at the time of decoupling).  In the hierarchical
CDM structure formation
models, the cool dissipationless gas of dark matter
first forms small structures that are the building blocks of 
successively larger structures.  These  eventually form the gravitational 
potential wells into which baryonic matter fall to form the observed galaxies.

The detection of this
dark matter has been a priority of the physics community over the past decade;
the astrophysical motivation for cold dark matter is not changed by the
recent discoveries.  The composition of this dark matter is
unknown, but astrophysical data suggest that any weakly interacting
massive particle (WIMP) could provide a natural explanation.  
The argument for weakly interacting particles goes as follows 
\cite{jung-kamio-greist-95}:

Initially the universe was very hot and dense and all particle species
were in thermal equilibrium.  As the universe expanded, particle species
initially followed their equilibrium value until the interaction 
rate $\Gamma_A = n_\chi \langle\sigma_A v\rangle$ fell below the Hubble 
expansion rate $H$, resulting in decoupling or {\it freeze-out}.  
For stable particles, the number density per co-moving volume 
after freeze-out is constant.
From this freeze-out condition, 
$\Gamma =n_\chi\langle\sigma_A v\rangle = H$,
it is apparent that the larger the value of the thermally averaged
product of the annihilation cross section and the relative velocity
$\langle\sigma_A v\rangle$,
the longer the particle stays in equilibrium, and the more it is 
Boltzmann suppressed before freeze-out.

A quantitative solution of the Boltzmann equation yields the
present mass density in units of the critical density:
\begin{equation}
\label{eq:relic}
\Omega_\chi h^2 \approx {3\times10^{-27}
{\rm cm^3\ sec^{-1}} \over \langle\sigma_Av\rangle }\quad .
\end{equation}
For a particle of mass $m_\chi$ with weak-scale interactions,
\begin{equation}
   \langle\sigma_A v \rangle 
   \sim \left({m_\chi \over 100\ {\rm GeV}}\right)^2 
   10^{-25}{\rm cm^3\ sec^{-1}}\quad ,
\end{equation}
which is very close to the value required to provide $\Omega = 0.3$.
This means that if there exists a new stable massive particle with weak-scale
interactions (i.e., a WIMP), it is a natural candidate for the dark matter.

Perhaps the best motivated of these particles is the neutralino
\cite{ellis-84}, 
the lightest, stable supersymmetric particle predicted by the 
supersymmetric extension of the standard model (see Jungman, Kamionkowski
and Greist \cite{jung-kamio-greist-95}
for a comprehensive discussion of neutralino dark matter
and the prospects for detection).  
Calculations of the relic abundance and mass of the neutralino suggest 
that they could be the dark matter if their mass is below a few TeV.
  
At this meeting J. Ellis and others presented updated calculations
of relic 
abundances taking into account co-annihilations and resonances in
the annihilation cross-section in the early universe.  These effects
destroy the simple one-to-one
relationship between the two-photon annihilation cross section
in the present universe and the total annihilation cross section at the
time of decoupling, and alter our expectations about the cosmologically
viable range of parameter space.
Without fine tuning, the likely mass range of the neutralino 
is $\sim$100~GeV to $\sim$1~TeV. 
However, Higgsino-like neutralinos with masses up to 8~TeV are possible 
without violating the upper limit $\Omega_{\rm CDM} < 0.3$.  
(These models are considered somewhat distasteful as they require 
fine-tuning and are viewed to be less satisfactory in solving the 
hierarchy problem.)
Of course if the anomalous muon magnetic moment $g-2$ result stands,
then the limit changes.   If the DAMA observation of an annual modulation
in the detector signal is valid, the neutralino
may have already been discovered (see the report of P4.6 in these 
proceedings).  The implications of 
the $g-2$ measurement are made somewhat murky by 
theoretical uncertainties in QCD, and the DAMA direct detection results
may be undermined by systematic errors in understanding background
rates in the detectors.  

If the dark matter in the galactic halo is comprised of cold, 
weakly interacting particles (e.g., neutralinos) then these particles 
should annihilate with each other giving a potentially observable 
signal in a number of different annihilation channels.  
The annihilation rate is proportional to the thermally averaged product
of the annihilation cross section and relative velocity, 
and to the square of the dark matter halo density.

Annihilation will result in direct or cascade production of protons, 
antiprotons, electrons, positrons, neutrinos and gamma rays that 
can stand out above the corresponding cosmic-ray backgrounds.  
Such {\it indirect} detection techniques are sensitive
to astrophysical uncertainties that enter into an estimation of 
the halo density and the backgrounds, but nonetheless can complement 
direct and accelerator searches.  

For a reasonable fraction of parameter space (covering varying
values for the neutralino mass and annihilation cross-section)
the local flux of antiprotons and positrons
can give a signal detectable by the current generation of cosmic-ray 
experiments.  The only astrophysical uncertainty that enters into 
the calculated flux comes from the relatively well established
local halo density.  Unfortunately, the 
signal has proven very difficult to discriminate from other 
cosmic-ray backgrounds, especially in the case of the anti-proton
flux~\cite{berg-edsjo-ullio-99}.  The recent measurement
of the cosmic-ray positron fraction by the HEAT experiment 
\cite{barwick-97, coutu-01},
may reveal a hint of an excess from the
annihilation signal, but uncertainties in backgrounds again make
it very difficult to draw a definitive conclusion \cite{baltz-01}.

A high energy neutrino signal could also be observed and 
has provided an impetus for proposals to build $\sim$km$^3$ 
neutrino detectors.  The gravitational field of the Earth and Sun
are expected to enhance the local density of neutralinos in the core 
of these objects, thereby substantially increasing the annihilation rate.
The emergent high energy neutrinos would offer a tell-tail signature,
since no standard astrophysical process could explain GeV to TeV neutrinos
emerging from the centers of such garden-variety celestial bodies! 
The predicted signal depends on the local halo density and the capture 
cross-section, but is thought to be reasonably well understood.  
Future detectors such as ICECUBE will be sensitive to a 
sizable fraction of the particle physics parameter space.  

Annihilation to charged particles will result in cascade emission 
of continuum gamma rays, and direct annihilation to two gamma rays 
(or a $Z$ and a gamma) will result in two nearly monochromatic lines 
at roughly the neutralino mass.  While the monochromatic line offers 
a smoking-gun signature for the dark matter, an observable flux
is generally only expected in the direction of the galactic center
where the dark matter
density peaks.  The predicted flux is much more sensitive 
to the detailed astrophysical model for the halo than the positron, antiproton
or neutrino measurements.
 
Recent N-body simulations of cold-dark-matter
structure formation indicate the possibility 
of cusps in the central density profile.  This has resulted in considerable
optimism, since if true it would mean that the gamma-ray detection 
channel could potentially provide access to essentially the {\it entire}
parameter space, a smoking-gun signature in the form of a narrow 
annihilation line that could not be confused with any astrophysical 
background, a measurement of the neutralino mass, and a measurement 
of the dark matter halo profile. 

The detectability of a gamma-ray signal from neutralino annihilation
is extremely sensitive to
the density profile of dark matter within the galaxy~\cite{berg-ulli-buck-98}.
Numerical simulations of the growth of structure in the universe suggest that
the mass density of dark matter halos may increase as a power law towards the
centers of galaxies. The existence of these density cusps is
controversial: there is some observational evidence against such dark
matter cusps~\cite{cote-etal-00,debl-etal-01,blai-etal-01}, other evidence
may be consistent with them~\cite{vdbo-swat-01,font-nava-01}.
If there is a density cusp at the galactic center
the neutralino annihilation rate may be large enough so that the
resulting rays would be detectable above the background by atmospheric
Cherenkov detectors and GLAST\cite{berg-ulli-buck-98}.

In the halo, the number of gamma rays produced
per unit volume per unit time is given by the source function
\begin{equation}
q_\gamma = 2 \langle\sigma_{\chi\chi\rightarrow\gamma\gamma}
  v\rangle n_\chi^2 \quad .
\end{equation}
The observed flux is obtained by integrating this source function
along the line of sight. 
If we observe at an angle $\theta$ with respect to the direction of
the galactic center, into an infinitesimal solid angle $d\Omega$,
then summing over the differential elements of volume $l^2dld\Omega$
along the line of sight gives
\begin{equation}
 {d\phi_\gamma\over d\Omega}(\theta) =
         \int_0^\infty q_\gamma(r) {1\over 4\pi l^2} l^2 dl\quad .
\end{equation}
This can be rewritten as
\begin{equation}
\label{eq:flux}
 {d\phi_\gamma\over d\Omega}(\theta) = 3.7\times 10^{-13}
\left({\langle\sigma_{\chi\chi\rightarrow\gamma\gamma}
  v\rangle \over 10^{-29} {\rm cm}^3{\rm s}^{-1}} \right)
\left({100\ {\rm GeV}\over m_\chi}\right)^2 I(\theta)\ \
{\rm cm}^{-2}{\rm sec}^{-1}{\rm sr}^{-1}\quad ,
\end{equation}
where the dimensionless line of sight integral $I(\theta)$ has been
normalized using
our galactocentric distance $R_\odot \approx 8.5$~kpc and the
local halo density $\rho_\odot \approx 0.3$~GeV~cm$^{-3}$ (as determined
from measurements of the galactic rotation curve) and is given by:
\begin{equation}
I(\theta) = (\rho_\odot^2 R_\odot)^{-1} \int_0^\infty
\rho^2\left(r(l,\theta)\right) dl
\end{equation}
where
\begin{equation}
r(l,\theta) = \sqrt{l^2+R_\odot^2-2l R_\odot \cos\theta} \quad .
\end{equation}

\begin{figure}[t!]
\begin{center}
\includegraphics[height=0.1\textheight]{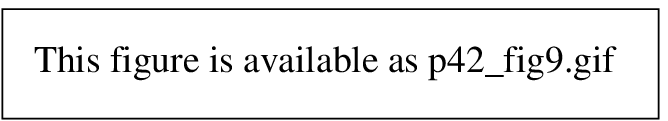}
\end{center}
\caption{(a) Value of the line of sight integral $I(\theta)$
for four different halo models (b)
relative signal to background ratios for these models as a
function of the angular acceptance interval given by
the solid angle $\omega$ about the direction $\theta= 0^\circ$.
Also shown are three models where the cusp has a constant
density core.
}
\label{fig:halosn}
\end{figure}

Figure~\ref{fig:halosn} shows the line of
sight integral $I(\theta)$ calculated for a number of different
halo models, variants of typical centrally-peaked profiles (without
spikes) as well as models
with constant density cores of varying radius (assuming
that the central halo was disrupted \cite{ulli-zhao-kamio-01}. 
For comparison, we show the more traditional (but perhaps
less well motivated)
isothermal halo that lacks the central density enhancement required
to give an appreciable annihilation signal.
Also shown in Figure~\ref{fig:halosn} is the relative
signal to background ratio ($S/\sqrt{B}$) for observations centered
on the galactic
center as a function of the angular acceptance window given
by the solid angle $\omega$.  This assumes that the observations
are dominated by an isotropic background so that
$\sqrt{B} \propto \omega^{1/2}$.
All halo models have been
renormalized so that the total mass inside the solar radius $R_\odot$ is the
same as that of the standard isothermal halo, thus matching the observed
galactic rotation curve at large distances from the galactic center.
The signal is obtained by
integrating $I(\theta)$ over a region of solid angle $\omega$ centered
on the galactic center. Thus
\begin{equation}
S/\sqrt{B} = \omega^{-1/2} \int_{\omega} I(\theta)\, 2\pi\,
\theta\, d\theta\quad .
\end{equation}

\begin{figure}
\begin{center}
\includegraphics[width=0.45\textwidth]{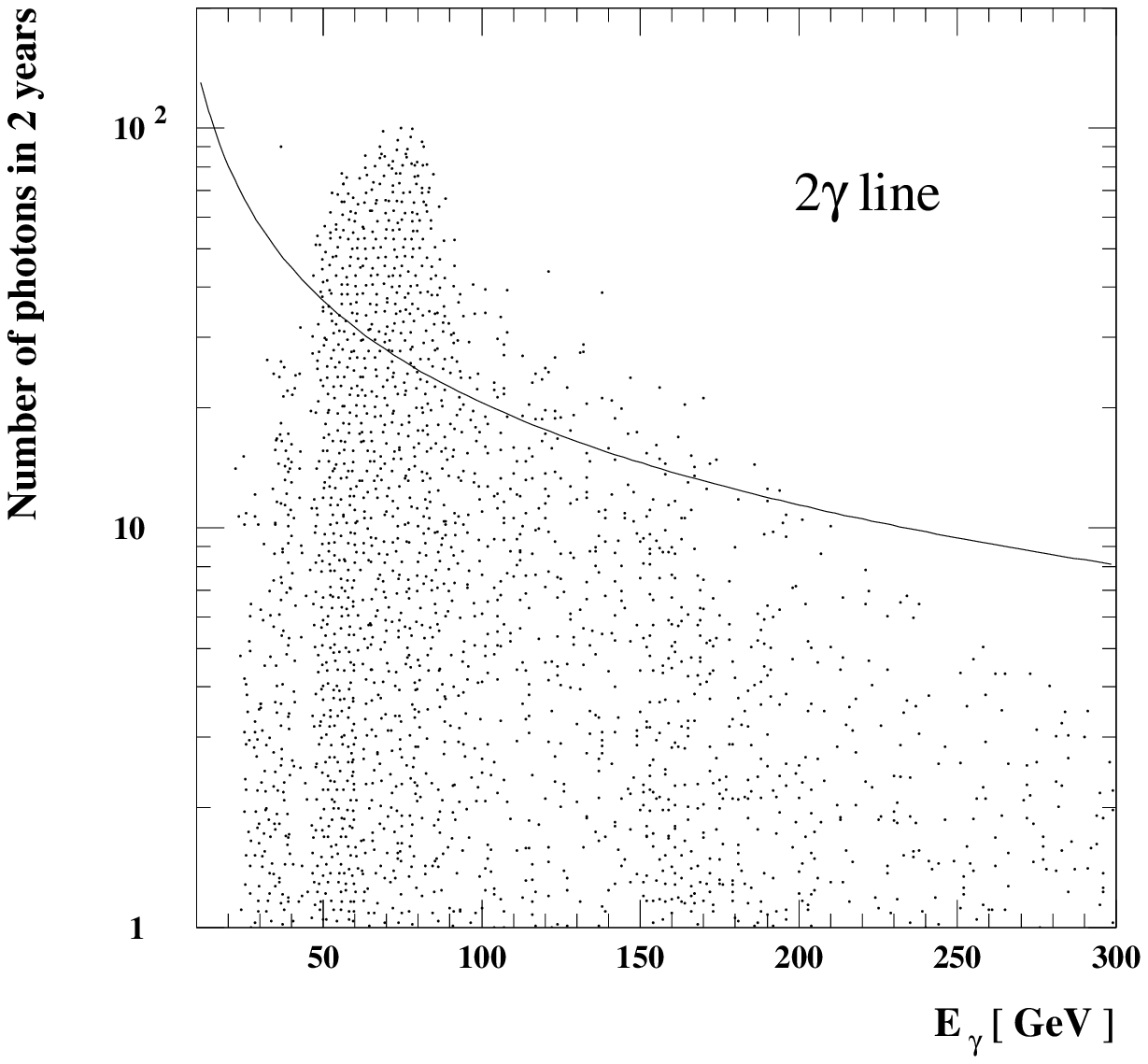}
\includegraphics[width=0.45\textwidth]{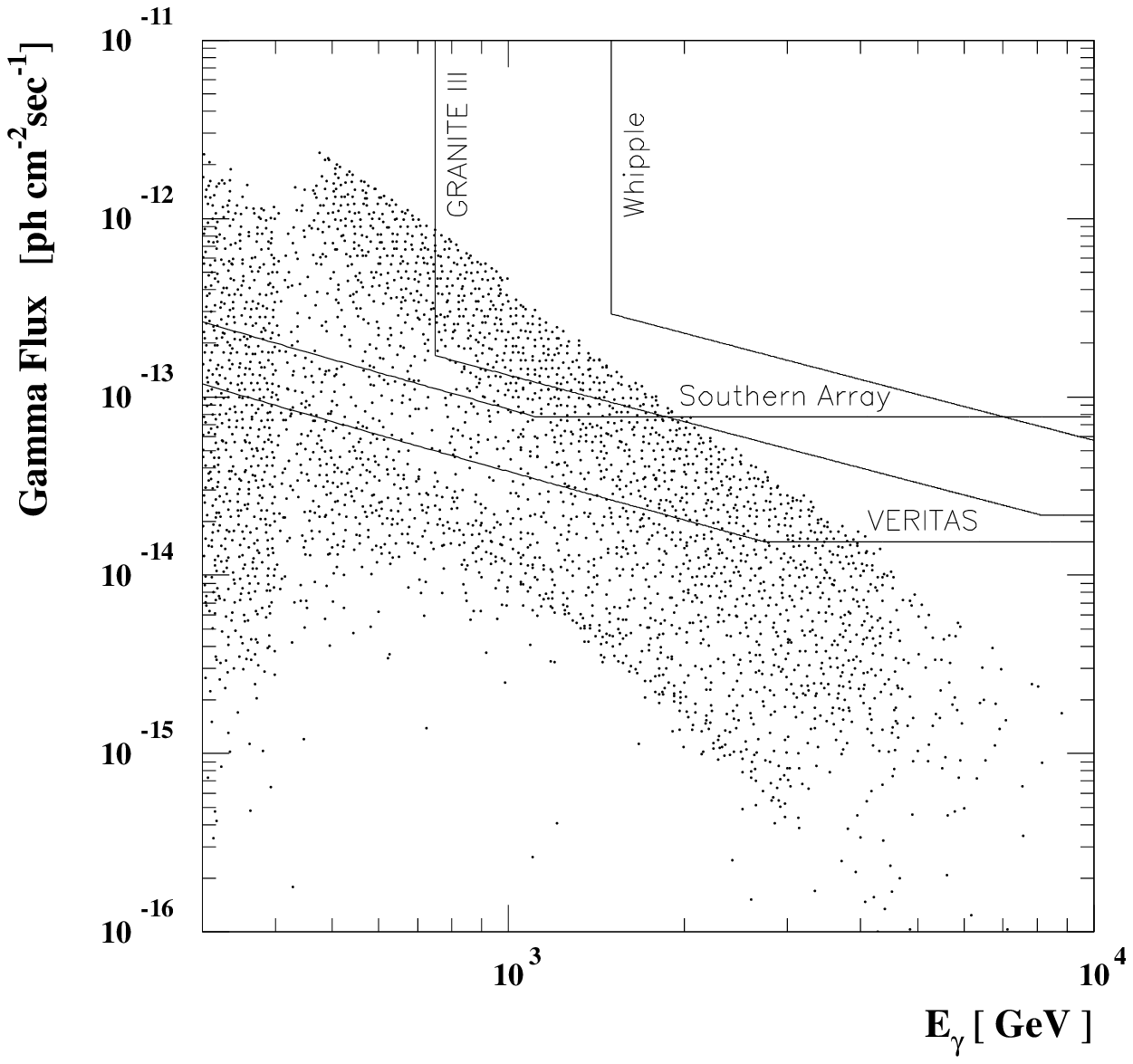}%
\end{center}
\caption{Predicted gamma-ray line flux for allowed supersymmetric parameter
space assuming a NFW halo \cite{berg-ulli-buck-98}.  {\sl Left} GLAST sensitivity
based on preliminary specifications of the wide-field calorimeter mode
\cite{berg-ulli-buck-98}.  {\sl Right:}  The expected
sensitivity for a five year exposure at 
large zenith angles with VERITAS or HESS, and
small zenith angle observations (made for the same duration) from the
southern hemisphere.}
\label{fig:bub}
\end{figure}

\begin{figure}
\begin{center}
\includegraphics[height=0.1\textheight]{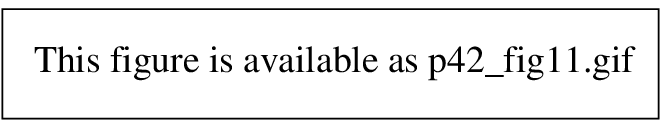}%
\end{center}
\caption{If the observed excess in the positron fraction measured with the
HEAT experiment is correct, then GLAST and ACTs might be expected to
verify the signal.  Above is shown the predicted gamma-ray line flux, and
sensitivity of one of the ACT detectors currently under construction.
The flux from both high galactic latitude and from the galactic center
is indicated assuming a conservative halo model, and showing only models
that pass constraints from fitting the positron data
 (Adapted from \cite{baltz-01}).  
  }
\label{fig:positron}
\end{figure}

Combining this astrophysical input with the particle physics calculation
of the annihilation cross-section for different supersymmetric models
one obtains the results shown in Fig.~\ref{fig:bub} \cite{berg-ulli-buck-98}.
Here the predicted
annihilation flux for the NFW halo are shown as a function the neutralino
mass.  Only models satisfying the cosmological and accelerator constraints
are shown.  Also shown is the sensitivity of GLAST and ground based 
instruments to the line flux.  Since
the resulting signal to noise ratio shows a strong dependence on the
detailed halo model, a comparison of the SNR for the NFW halo
and for other
halo models (as shown in Fig.~\ref{fig:halosn}) can be used to slide the
sensitivity curves up or down with respect to the model points; for some
halo models with relatively steep central cusps almost the entire parameter
space is accessible.  If the observed bump in the HEAT positron spectrum
is really an indication of neutralino annihilation, then the allowed parameter
space is as shown in Fig.~\ref{fig:positron} \cite{baltz-01}.  Future ground
arrays such as MAGIC, VERITAS and HESS would be sensitive to almost the
entire parameter space for halo models with a modest cusp.

The issue of the existence of WIMP cusps is exacerbated by an additional,
possibly strong, concentration of dark matter around the black hole at the
galactic center~\cite{gond-00,gond-silk-99}. 
(Traditionally, this additional concentration
follows a power law in density and would be called a cusp, but, to
distinguish it from the halo density cusp of N-body simulations, it has
been referred to as a spike.
We define any power law density profile with a spectrum
steeper than $ \sim r^{-2} $ around the black hole as a steep spike, and 
any shallower power law profile as a shallow spike.)
The spike could originate during the gradual (adiabatic) growth
of the central black hole as dark
matter is trapped and the gravitational
potential gets deeper. A similar phenomenon occurs with stars (which also
act as a dissipationless gas)
and is believed to be the cause of the observed stellar
concentrations around the central black holes of elliptical galaxies and bulges
of spiral galaxies.

The dark matter spike at the galactic center may be shallow or steep according
to the phase space density of dark matter at the time the black hole
formed~\cite{gond-silk-99}, depending on whether the center 
of dark matter distribution
and black hole were coincident, and depending on whether the black hole
was formed adiabatically or through impulsive mergers \cite{ulli-zhao-kamio-01}.
At this meeting Merritt described simulations of baryonic and dark matter
structure formation at the centers of galaxies with massive black holes.
His simulations show that the stellar and dark matter spikes could be destroyed
by mergers, explaining why stellar spikes are only observed in a subset
of galaxies. 

\begin{figure}
\vspace*{-0.8in}
\begin{center}
\includegraphics[width=0.60\textwidth]{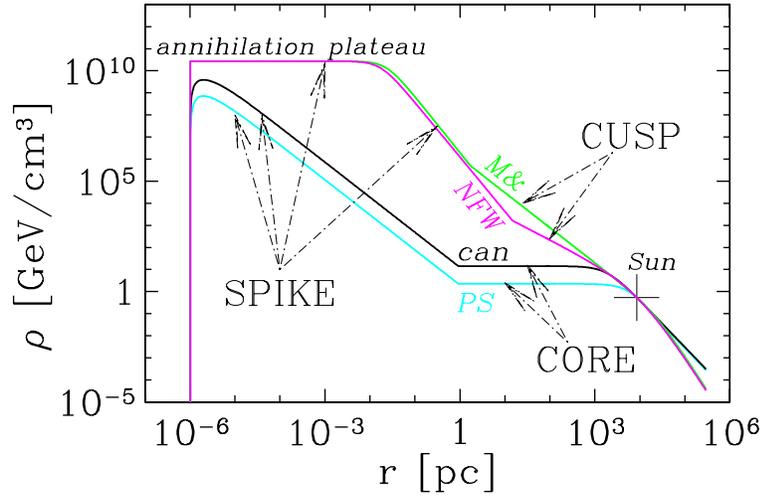}%
\end{center}
\caption{Dark matter density as a function of distance from the galactic
  center, showing spikes, cusps, and cores. Four density models are shown:
  canonical (``can''), Persic--Salucci (``PS''), Navarro--Frenk--White
  (``NFW'') and Moore-et-al.\ (``M\&''). The models are normalized to the same
  rotation velocity and the same density at the Sun's position (marked by a
  cross).}
\label{fig:spike}
\end{figure}

Gondolo and Silk
\cite{gond-silk-99} considered the case of adiabatic growth of a black
hole coincident with the center of the dark matter distribution.
If there was a high density of slow dark matter particles, and
a black hole would form gradually at the center of such a halo,
the spike would be steep.
If the initial distribution was isothermal,
the dark matter spike would be shallow, leading to a dark halo
with a central core.  Figure~\ref{fig:spike} illustrates steep and shallow
spikes for four models of a WIMP galactic halo. Two of the models shown, the
canonical model~\cite{bahc-sone-80} (marked ``can'') and the model by Persic,
Salucci, and Stel~\cite{pers-salu-96} (marked ``PS''), have a core. The other
two models, that by Navarro, Frenk, and White~\cite{nava-fren-96} (marked
``NFW'') and that by Moore et al.~\cite{moor-etal-98} (marked ``M\&''), have a
cusp.  Both types of models have a central spike, models with a cusp having a
steeper spike.  The density in the spike follows essentially a power law,
except when it becomes so high that the WIMP annihilation timescale is
comparable to the age of the black hole. In this case, the density remains
constant (``annihilation plateau'').  The spike's innermost radius occurs where
the WIMPs are captured by the black hole.

The steepness of the spike is crucial when considering the annihilation of
WIMPs in the spike: while a shallow spike is undetectable, a steep spike may
give rise to a detectable signal in high-energy neutrinos and
gamma-rays~\cite{gond-silk-99}. Furthermore, with a steep spike, electrons and
positrons generated in WIMP annihilation may produce synchrotron emission well
in excess of the observations~\cite{gond-00}, ruling out WIMP dark matter
at the galactic center.

Gondolo argues that 
the most natural way to avoid a steep spike is not to have a halo cusp in the
first place, in line with the observations against the existence of halo
cusps~\cite{cote-etal-00,debl-etal-01,blai-etal-01}. Indeed a series of
suggestions to eliminate halo cusps have been proposed, some of which do
not dismiss WIMP dark matter~\cite{kami-lidd-00,binn-etal-01,wein-katz-01}.
Without a halo cusp, however, the prospect of detecting WIMP dark matter
through gamma-ray emission is bleak.  

However, this conclusion seems to have some significant caveats.
It is possible to have
a halo cusp without a steep central spike if the
black hole did not form at the center of the dark halo or if the present black
hole is the result of merging two or more previous black
holes~\cite{ulli-zhao-kamio-01,milo-etal-01}. It is still an open question if
these scenarios are compatible with the observed existence of a steep {\it
 stellar} spike around the black hole at the galactic center~\cite{alex-99}.
Until the theoretical uncertainties
are resolved any non-detections of WIMP annihilation lines give poor
constraints on the existence of WIMPs.  However, the possibility of 
large density enhancements at the galactic center provide the potential
for discovery, and mean that gamma-ray
observations may provide one of the few methods for which the entire
allowed parameter space is accessible.
Of course a detection of gamma rays from WIMP annihilations would be
enormously important since it could provide a smoking-gun signature for
nonbaryonic dark matter, provide indirect evidence
for supersymmetry, would provide a measurement of the dark matter mass, and
would also provide information about the dark matter profile that would
serve as a sensitive test of our theoretical understanding of structure
formation.

\begin{figure}[t!]
\vspace*{0.0in}
\begin{center}
\includegraphics[height=0.1\textheight]{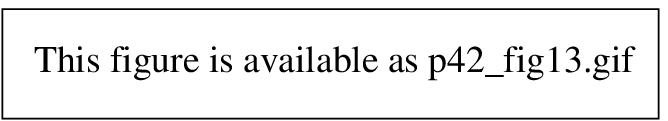}
\end{center}
\label{fig:egret}
\vspace*{0.2in}
\caption{EGRET observations of the galactic center region
from Mayer-Hasselwander et al. (1997).
(a) background-subtracted, smoothed $E_\gamma >1$GeV count map,
(b) 300~MeV -- 1~GeV map.}
\end{figure}

To date there have been some modest searches for the annihilation
line with only upper limits on the flux \cite{kosack-01}.
However, it is interesting to note that
the observed GeV flux at the galactic center is consistent
with the continuum flux predicted in some of the more optimistic
halo models.  While this does not constitute a detection and may well
be associated with some other gamma-ray source, there is
certainly room for the continuum component \cite{mayer-hasselwander-98}.
Clearly the detection of an annihilation line will be required to
obtain an unambiguous result.

\subsection{The Cosmic Infrared Background}

Gamma-rays also can be used as probes of the
the extragalactic background light (EBL), the relic starlight
that fills the universe.  The diffuse infrared background
light contains contributions from redshifted starlight,
starlight that has been scattered by dust,
and any of a number of processes in the history of the universe that
have pumped significant amounts of energy into the infrared (IR) to
energy band.  A large contribution to this primordial radiation field 
comes from the era of galaxy formation, when most of the star
formation occurred.
Direct measurement of the characteristics of this radiation are
difficult due to the dominance of local galactic IR sources. 

TeV gamma-ray astronomy provides a means to study the IR background
free of such complications by looking for modifications to high energy
gamma-ray spectra.  Gamma rays from distant AGN
interact with this background field via
the process $\gamma\gamma\rightarrow e^+ e^-$ \cite{Gould67,Stecker93}.
Since this is a resonant process, with a large cross section when the
center-of-mass energy of the two photons is equal to twice the electron mass,
the spectra of distant AGNs becomes distorted.  The quantitative effect on the 
AGN spectrum depends on the spectrum of the EBL.

\begin{figure}[t]
\begin{center}
\includegraphics[height=.47\textheight]{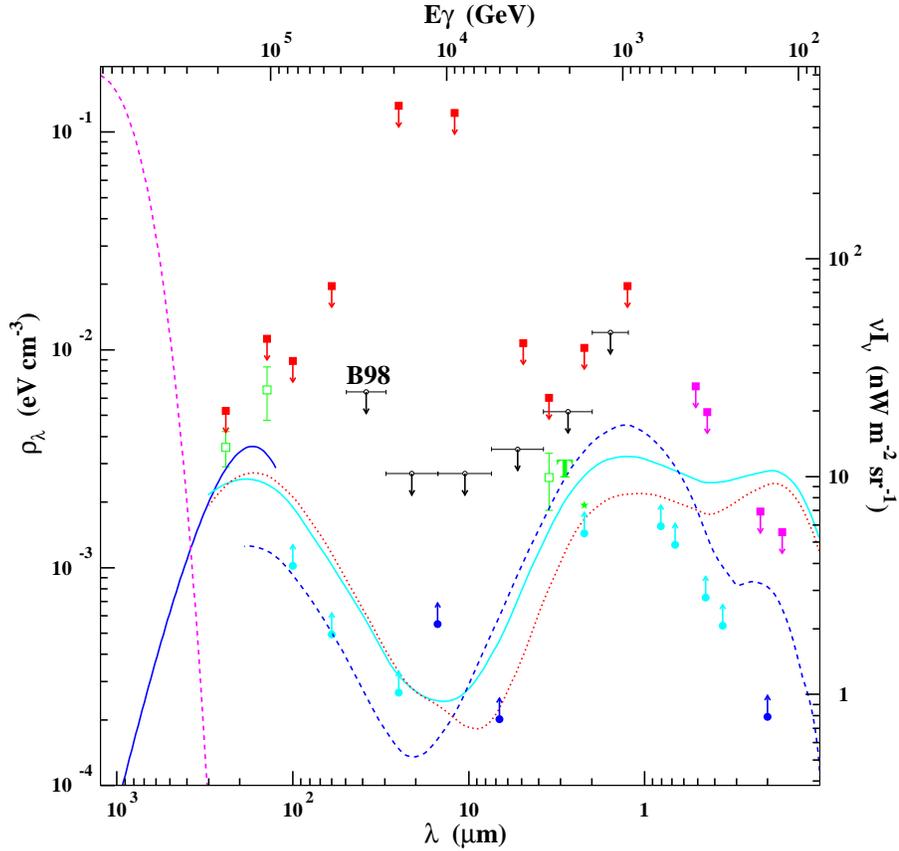}
\end{center}
\caption{Spectrum of the extragalactic background light with
initial {\sl measurements} by DIRBE, upper limits by
Whipple \protect\cite{Biller98} and lower limits from K-band
galaxy counts (adapted from \protect\cite{Vassiliev00}).
\label{fig:ebl}}
\end{figure}

Measurements of the spectra of the TeV sources
Mrk~421 and Mrk~501 have been used to set an upper limit
on the density of extragalactic background photons (under the assumption
that the intrinsic source spectra do not become harder or rise exponentially
above a few hundred GeV).  The resulting
limits are more than an order of magnitude more restrictive than
direct observations in the 0.025-0.3~eV regime \cite{Biller98} and are shown in Figure \ref{fig:ebl}.
Also shown in the figure are direct measurements by DIRBE and lower
limits come from K-band galaxy counts.  Model curves from
\cite{Primack99} show sensitivity to the history of star formation
(e.g., CHDM vs. SCDM) in the 1-10$\mu$m regime and
sensitivity to the initial mass function and re-radiation by dust
in the interval ($\lambda\lesssim$1~$\mu$m and $\lambda\gtrsim$10~$\mu$m).

Combining the numerous AGN spectra measured with GLAST with
the ground-based experiments one will obtain the detailed spectra
of a large number of sources required to disentangle the intrinsic
absorption from the intergalactic absorption, providing a measurement
of the EBL.  This is an important example where space and ground-based
instruments acting together are much more powerful
than either instrument acting alone. 
 To make this measurement in the infrared portion
of the spectrum, it is critical that the GLAST measurements of the largely
unabsorbed spectra at lower energies be complemented by measurements with
ground-based telescopes.  Above 50 to 100 GeV, the huge effective area of
ground-based experiments will be required to overcome the counting-statistics
limitation of satellite experiments
and measure the spectral cut-offs.  The overlap with GLAST
will be critical to calibrate the ground-based instruments.  Furthermore,
while the ground-based telescopes may make the most sensitive measurements
of spectral cutoffs of some sources, the GLAST spectral 
measurements over almost 3 orders of magnitude (from 100 MeV to 100 GeV) will
provide information about the source spectrum
essential for normalizing the VHE measurements and making reliable constraints.

\subsection{Primordial Black Holes}

Stephen Hawking first postulated that black holes might in fact not be
completely black \cite{hawk74}.  Hawking showed that a black hole has an
effective temperature that is related to the mass of the hole by
$T=10^{13}/M$ GeV (where $M$ is given in grams).
Since the emission is nearly thermal the luminosity of the black hole
is given by ${\cal{L}} \propto 1/M^{2}$. 
As the hole radiates energy, it's mass decreases, it gets hotter and
radiates more rapidly.
This runaway process leads to the eventual evaporation of the hole.
Black holes formed during the early universe with a mass $\sim10^{15}$
grams would be evaporating now ($t\approx10^{10}(M/10^{15})^3$ yr).
The evolution of the luminosity of a black
hole depends upon the number of degrees of freedom available.  As the
temperature rises and heavier particles can be created the luminosity
increases.  In principle one can probe the particle spectrum up to the
Planck scale by measuring the luminosity evolution of an evaporating
black hole.  There is further interest in
the Hawking radiation because it relies on the application of 
relativistic quantum field theory in the presence of the strong field
limit of gravity, a situation that could potentially be observed. 

To set the scale from fundamental physics, we note 
that the spontaneously broken gauge symmetry in the
electroweak sector of the standard model gets restored in a phase
transition or rapid crossover at a temperature near 100 GeV.  The 
fact that temperatures of the latter order of magnitude will never be 
achieved in a terrestrial experiment should motivate us to study the 
fate of microscopic black holes during the final few days of their 
lives when their temperatures have risen to 100 GeV and above. The fact 
that microscopic black holes have not yet been observed should not be 
viewed as a deterrent, but rather as a challenge for the new 
millennium.
Lacking a firm cosmological prediction for the relic abundance,
we address the question:
What would be the observable signal?

When the Hawking temperature exceeds a few hundred MeV we should expect 
that jets of quarks and gluons are emitted which then
fragment into hadrons, an idea first proposed by MacGibbon,
Webber and Carr \cite{macgibbon}.  
Subsequently Heckler \cite{heckler1,heckler2} argued that at sufficiently
high temperatures the emitted quarks and gluons are so densely packed
outside  of the event horizon that
they do not fragment into hadrons in a vacuum but in
something more akin to a quark-gluon plasma. 
He also argued that QED bremsstrahlung and pair production were 
sufficient to lead to a thermalized QED plasma when $T_H$ exceeded 45 
GeV. A more quantitative treatment of the particle interactions on a 
semi-classical level was carried out by Cline, Mostoslavsky and
Servant \cite{cline}, who solved the relativistic Boltzmann equation with QCD
and QED interactions in the relaxation-time approximation. They found 
that significant particle scattering would lead to a photosphere, though 
not perfect fluid flow.

Daghigh and Kapusta \cite{kapusta} applied relativistic viscous 
fluid equations to the 
problem when the Hawking temperature exceeds 100 GeV. They found that a 
self-consistent description emerges of an outgoing fluid or wind
from the black hole which is just marginally kept in local thermal
equilibrium, and that viscosity plays a crucial role in the dynamics.
A sizable fraction of the total luminosity from the 
photosphere (which can be many orders of magnitude greater in radius
than the Schwartzschild radius) is in the form of photons.  They calculated the 
instantaneous and time integrated photon flux.  Using the present upper 
limit on the rate density of microscopic black holes of about 1 pc$^{-3}$ 
yr$^{-1}$ in our neighborhood, they showed that the diffuse spectrum is 
unlikely to be detectable above other sources.  A much more promising 
route is the search for individual explosions.  Assuming no new physics 
beyond the standard model, a black hole with a temperature of 100 GeV 
will disappear within 5.4 days.  Between 5.4 days and the final 10 
$\mu$s the average photon energy will increase from 4 to 160 GeV. 
The signal would be a source which gets brighter over a period of several
days and then suddenly disappears.  This would be a very unusual event.
The maximum reach of a detector which has a threshold for photons of
$E_{\rm min}$ and an effective area of $A_{\rm det}$ is
\begin{equation}
d_{\rm max} \approx 150 \sqrt{\frac{A_{\rm det}}{1 \, {\rm km}^2}}
\left( \frac{10 \, {\rm GeV}}{E_{\rm min}}\right)^{3/2} \, \, {\rm pc} 
\, .
\end{equation}
The above formula is valid for $E_{th}>10$ GeV.
If we take the local rate density of explosions to be 0.4 pc$^{-3}$ 
yr$^{-1}$ (the current upper limit) then within 1 pc of Earth there 
would be $\sim 1.5$ explosions per year. These would be distributed 
isotropically in the sky.  Still, it suggests that the direct 
observation of exploding black holes may be feasible if their abundance
is near the inferred upper limit. 

\subsection{Constraints on Quantum Gravity and Large Extra Dimensions}

Theories of quantum gravity, though incomplete, often
seem to result in violations of some of the pillars of
modern physics including Lorentz invariance, and without some care,
even causality.  Most theories of the small-scale
structure of spacetime lead to dispersion relationships for light traveling
in vacuum.  The energy dependence of the velocity of light
from such relationships is not to be
confused with photon mass, which gives the opposite sign.  Instead, photons
with energies approaching some characteristic TeV to Planck scale are
retarded either through the exchange of a Planck scale particle present
in quantum gravity, or through distortions in the metric on small
distance scales \cite{amelino98}.  This violation of Lorentz invariance
may be manifested as an energy dependent velocity of light. 
For many theories of quantum gravity the velocity of light can be
parameterized as 
\begin{equation}
v \approx c(1-\zeta\frac{E}{E_{QG}}),
\end{equation}
where $E_{QG}$ is an energy scale related to quantum gravity and $\zeta$
is of order 1 and we have ignored terms of higher order in $E/E_{QG}$.  
Since $E_{QG}$ is expected to
be quite large (of order the Planck mass) there seems to be little hope
of measuring this effect
with earth-bound experiments.  To measure small velocity differences one
needs short pulses of very high-energy photons
traveling cosmological distances. 

\begin{figure}[t]
\begin{center}
\includegraphics[height=0.35\textheight]{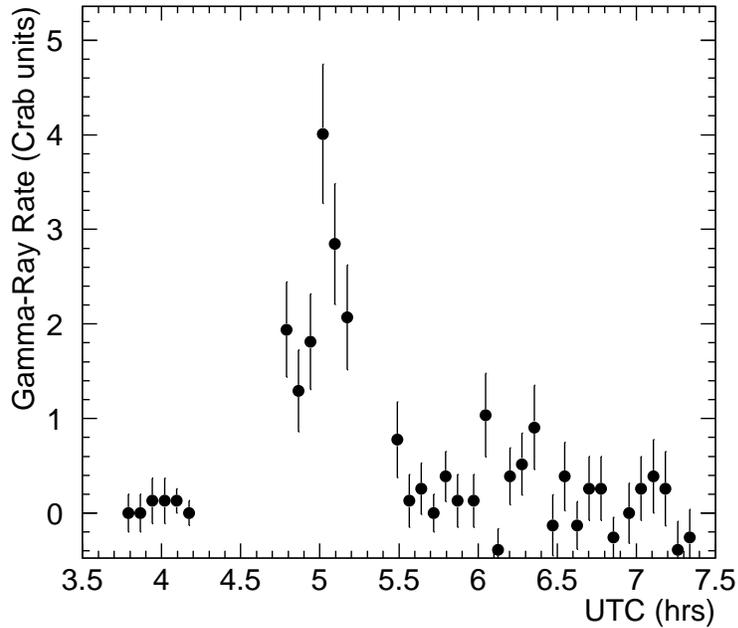}
\end{center}
\caption{Mrk 421 flare observed on May 15, 1996 (reprinted with permission
of {\sl Nature} 383, 26 September 1996, pp. 320-321; \copyright 1996 Nature
Publishing Group\protect\cite{Gaidos96}
\label{fig:flare15min}}
\end{figure}

The above equation leads to a measured time delay for photons of
energy $E_1$ and $E$ of 
\begin{equation}
\Delta T  \approx \zeta \frac{LE}{cE_{QG}}
\end{equation}
if $E$ is much larger than $E_1$.  Thus a figure of merit for such
measurements is $LE/\Delta t$.  There are two known sources with reasonable
figures of merit: gamma-ray
bursts (GRBs) and active galactic nuclei.  As discussed  above
GRBs are short intense bursts of gamma
rays that are visible across the visible universe.  Active galaxies, while
closer, still lie at relatively large
distances, and exhibit large short term flux variations.  For example,
Fig.~\ref{fig:flare15min} shows a half-hour flare of TeV photons from an AGN at
a distance of 100~Mpc (Mrk~421) observed with the
Whipple gamma-ray telescope.   These systems make ideal laboratories for
studying the constancy of the velocity of light. 
A difficulty in using GRBs to measure this dispersion is that there
are only a few bursts for which the redshift is known and the energy at which these bursts have
been observed is typically <100 keV.  To date the best limit on
quantum gravity has been set by the Whipple
collaboration \cite{biller}.  In May 1996 the active galaxy Mrk421 ($z=0.031$)
emitted a flare that lasted for 280 seconds.  By examining the arrival time
of low (<1 TeV) and high (>2 TeV) energy data they established a lower limit
to $E_{QG}$ of $6\times10^{16}$ GeV.  However, this limit assumes that the
higher energy photons were not emitted substantially earlier than
the lower energy photons.  While such an effect is not predicted,
there is at present
no way of ensuring that an effect due to quantum gravity wasn't partially
masked by details of the source; this
points to the necessity of understanding
the astrophysics of the source to make robust constraints.
 More recently Ellis et al. \cite{ellis} have used
a sample of GRBs from different redshifts to set a lower limit
$E_{QG} > 10^{15}$ GeV.  Observations of rapid high energy
flares from distant AGNs and GRBs 
with instruments such as GLAST, Milagro, and VERITAS
should improve the current limits
by several orders of magnitude and may even reach the Planck scale.

As with all astrophysical measurements it will be crucial to
understand the sources of the photons.  In this instance,
one must be sure that the higher energy photons were not emitted
earlier than the lower photons on timescales of seconds.  While this
 seems to be a formidable task, by making measurements of identical
(or similar sources) at different redshifts (GRBs), and by measuring
the same effect in a different type of source (AGNs), one 
should be able to make an unambiguous measurement of the effect,
even without completely understanding all of the complications of the
emission mechanism at the source.

\section{Techniques of Gamma Ray Astronomy}

The flux from all known sources of gamma rays falls rapidly
with energy.  For example the flux of photons from the Crab nebula is
$\sim 0.01\gamma's m^{-2} s^{-1}$
above 100 MeV but only $\sim 10^{-7}\gamma's m^{-2} s^{-1}$ above 1 TeV.  
The earth's atmosphere is opaque to high-energy photons (above the
ultraviolet), but higher energy photons generate cascades in the
atmosphere (extensive air showers), which are detectable from the ground.
At lower energies ($\sim$100 GeV) the Cherenkov radiation
generated by the electromagnetic particles in the shower can be detected
by air Cherenkov telescopes (ACTs), and at higher energies
sufficient numbers of particles survive to the ground to be directly
detected by extensive air shower (EAS) arrays.  At lower energies
direct measurements with space-based instruments have been used.
Over the past decades great progress has been made in extending the
energy reach of space-based instruments, with GLAST expected
to have sensitivity beyond 100 GeV, and lowering the energy threshold of
ground-based instruments.  A decade ago the Whipple ACT had an energy
threshold of $\sim$450 GeV, today experiments utilizing solar power plants,
such as STACEE and CELESTE, have achieved energy thresholds
below 100 GeV.  As for detectors that directly detect the particles on
the ground, a decade ago instruments
had achieved energy thresholds of $\sim$100 TeV.  Today, the Milagro
EAS array has an energy threshold below 1 TeV.  

The other critical feature of gamma-ray telescopes is their ability to
reject the cosmic-ray background, which is orders of magnitude larger
than the observed gamma-ray signals.  The space-based instruments can
be surrounded by anticoincindence shield and track the electron
positron pairs, providing a straightforward and powerful means
of rejecting the cosmic-ray background.  For ground-based instruments,
the equivalent of an anticoincidence shield does not exist and a detailed
understanding of the development of hadronic and electromagnetic showers
is used to reject the background.  Progress has been slow, but steady
improvements have been made.  The first generation of ACTs and EAS arrays
did not utilize any background rejection and did not convincingly 
detect any cosmic sources of gamma rays.  It was the landmark
development of the ``imaging'' technique, originally proposed by Weekes
and Turver \cite{weekes-img} that revolutionized the field of 
ground-based gamma-ray astronomy.  Modern ACTs utilizing this technique
can remove over 99.7\% of the cosmic-ray background yielding
unprecedented sensitivity in this energy range.  EAS arrays used
shielded detectors to detect the penetrating component (muons and hadrons)
of hadronic air showers.  While this technique has been used
for several decades, Milagro is the first instrument to utilize this
method in at low primary energies,
where one expects to observe sources of gamma rays.  

In this section we give a brief history of the 
various instrumentation used in gamma-ray astronomy,
discuss the next generation instruments that are currently
 under construction, and give a 
recommendation for the future development of the field.
  In the field of particle physics, new discoveries have been
dependent upon the construction of new machines, capable of
 attaining ever-higher energies.  The
situation is similar in astronomy and in gamma-ray astronomy
 in particular.  We have seen how the development
of new techniques has lead to new discoveries and insights into
 the universe.  It is clear that to continue
along the path of discovery we must develop new instruments, capable of
 seeing further into the cosmos (low energy
threshold) , and observing the entire sky with much greater
 sensitivity, to detect and study the
transient phenomena that pervade the TeV universe.

\subsection{Space-Based Experiments: EGRET and GLAST}

Above a few MeV, the gamma-nucleus cross section is dominated by pair
conversion, which dictates that instruments in this energy range be designed to
detect and reconstruct the electron positron pair direction and total
energy.

The useful energy range of such a ``pair telescope'' is determined by the
pair-production cross section (at the low end) and counting statistics
(at the high-energy end).  Although the threshold for pair creation
is 1 MeV, the cross
section is not significant until ~50 MeV, and the threshold for meaningful
detection is usually quoted as 20 MeV. The angular resolution is very poor at
this energy, however, with 100 MeV being a more practical lower limit for the
study of point sources.

The highest energy is limited by flux. With an effective cross sectional area of
$\sim$1 m$^2$, the rate of photons from the Crab nebula above 300 GeV
is $10^{-6}$ m$^{-2}$ s$^{-1}$. For a year's direct observation in a
detector with 1 m$^2$
effective area, the largest practical area for a satellite, this corresponds to
about 25 photons per mission, just enough for a ``5 $\sigma$" detection.

The three requirements of photon conversion, reconstruction of the pair or
subsequent shower, and measurement of the energy, lead to a configuration shared
by EGRET and the proposed GLAST: a front section with thin converters
interleaved with tracking system; followed by a calorimeter section.

The limiting sensitivity of $10^{-6}$ m$^{-2}$ s$^{-1}$ should
be compared with the
rate for cosmic rays and albedo protons and electrons to intersect the
detector, which is around 5 kHz, or some nine orders of magnitude larger.
The maximum rate at which data can be transmitted to the ground dictates
the efficiency of the hardware trigger.  For an average telemetry rate of
about 50 Hz,  the instrument have a
trigger that rejects 99\% of incoming cosmic rays with high efficiency for
gammas. This requires a third system, namely an outer shield to detect incoming
charged particles.

\subsubsection{EGRET}

The Compton Gamma Ray Observatory, including EGRET, was launched in 1991.  The
EGRET tracking system used 1 mm pitch wire spark chambers with core readout.
This technology had several important consequences, including 100 ms deadtime
for recharging, and a need for a very strict trigger to limit use of the
consumable spark chamber gas and ensure that the trigger rate was consistent
with the bandwidth for transmission to ground.

The EGRET trigger had two components: a requirement that there be no count in a
monolithic anticoincidence shield surrounding the instrument, and coincidence
between two planes of counters between the tracking and calorimeter sections,
separated by 30 cm so that the direction could be determined by time of flight.

The EGRET trigger design produced an additional constraint on the maximum
gamma-ray energy beyond the simple statistics limit.  Since the
anticoincidence could be triggered by ``back splash" from particles escaping
from
the calorimeter (a set of NaI crystals), high energy events could produce
a {\it self veto}. This self-veto limited its high energy
to about 10 GeV. The need for the time of flight trigger resulted in
lengthening the instrument, which in turn limited its field of view.  

\subsubsection{GLAST}

GLAST is scheduled for launch in early 2005. The payload is primarily a pair
conversion telescope, (Large Area Telescope, or LAT) with a small gamma-ray
burst monitor. The design has the same elements as EGRET, but has significantly
better performance as a result of modern solid-state technology.

The most important difference is that the tracking section uses silicon strips
detectors with 200 {$\mu$}m pitch. This allows dramatic improvements of the
deadtime, aspect ratio and therefore angular acceptance, angular resolution at
high energy, and high energy acceptance.

Since the tracking system can be read out in a few {$\mu$}s,
and doing so involves
no consumables, there is no penalty for a loose first-level trigger.
The data can then be
analysed and filtered on board before it is transmitted to the ground.
Thus it is feasible to trigger on
all incoming particles, with rates up to 10 KHz, without using the
anticoincidence system in the fast trigger. Like the anticoincidence system of
EGRET, GLAST is surrounded by scintillator, but is is composed of an array of
individual counters rather than a single monolithic dome. Thus the filtering can
be selective, in two respects not possible with EGRET: first, if a track
candidate determined by the rough on-board track recognition software does not
extrapolate to a counter with a signal, it is not vetoed. Second, if the total
energy deposited in the calorimeter section is large, the presence of
anticoincidence signals will be ignored, avoiding the self-veto possibility.

The resulting increase in the dynamic range of GLAST, improved background
rejection with it's higher resolution tracker, substantially wider field
of view and low dead-time detectors results in more than
an order of magnitude improvement in sensitivity compared with EGRET.  

\subsection{Extensive Air Showers}

\begin{figure}[htb]
\begin{center}
\includegraphics[width=12cm]{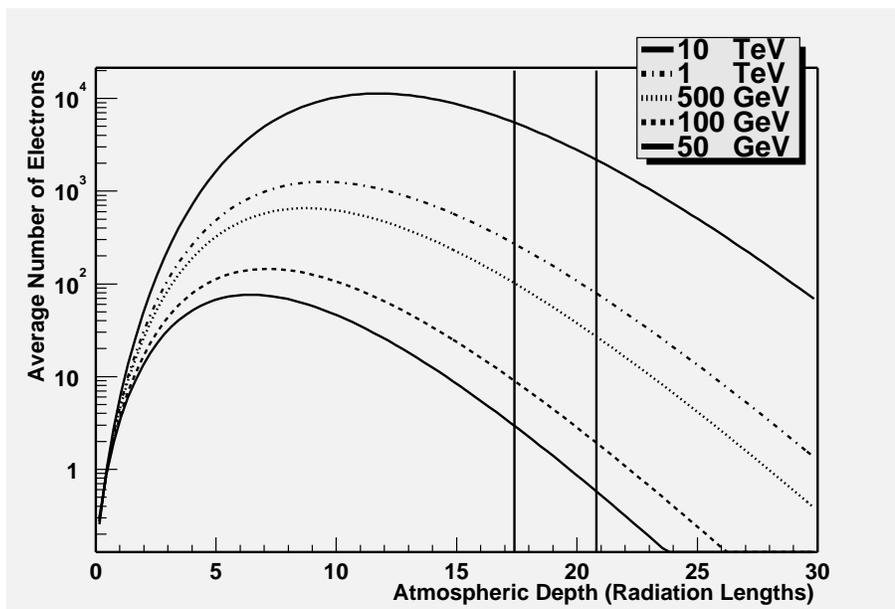}
\end{center}
\caption{Shower development curves as given by an approximate
semi-analytical expression. 
The curves show the average number of electrons (and positrons)
as a function of atmospheric depth for various energy 
primary gamma rays.  The vertical lines are drawn at 4000~m asl
 and at 2600~m asl.}
\label{fig:approxb}
\end{figure}

When a high energy gamma ray enters the earth's atmosphere it loses
energy by creating an electron
positron pair.  These particles then lose energy via bremsstrahlung,
producing more high energy photons.  Until the average energy per
particle reaches 80 MeV, particle creation processes dominate the energy loss
mechanisms of electromagnetic particles.    
When the average particle energy reaches 80 MeV, ionization
losses dominate and the number of particles in the shower begins to
decrease; this point in the development
of the air shower is called shower maximum.  The particles travel in
a rough pancake, that has a $\sim$100~m  
radius and $\sim 1$~m thickness when it reaches the ground. 
Figure \ref{fig:approxb} shows the longitudinal development of
extensive air showers of various energies.  These curves show the average 
number of electrons in an extensive air shower as a function of
atmospheric depth.  At ground level 
high-energy photons outnumber electrons by a factor of $\sim$4.
The development of a single air shower is dominated by fluctuations
in the development of the shower.  The dominant fluctuation being
the depth of the first interaction.     

\subsection{Air Cherenkov Telescopes}

At present, the most successful ground-based
technique employs imaging atmospheric Cherenkov telescopes (IACTs)
in which a large optical reflector and a pixelated
fast photomultiplier tube (PMT) camera is used to record the images of
electromagnetic cascades.  The imaging information can be used to
efficiently reject the large background of cosmic-ray showers by making
use of the characteristic differences of the hadronic and purely
electromagnetic showers (see Fig.~\ref{fig:events}).  The fast time
structure of the Cherenkov
pulses ($\sim$ a few nsec) is exploited to detect the
faint flashes above the Poisson fluctuations of the night sky background
light.  The power of this technique comes from the fact that the effective
area is much larger than the geometric area of the instrument, and is
given instead by the size of the Cherenkov light pool at ground level.
Depending
on the zenith angle (and distance to shower maximum) this gives an effective
area from 10,000~m$^2$ to 0.1~km$^2$.  The total amount of Cherenkov
light detected provides a reasonably good calorimetric measurement
giving an energy resolution of 25\% to 40\% for a single telescope.
While a single telescope provides hadronic rejection with an efficiency
of $\sim$99.7\%, arrays of telescopes can do substantially
better by using stereoscopic imaging to better discriminate hadronic
showers.  These instruments also provide better angular resolution,
more precise energy measurements, and a level of redundancy critical
to quantify systematic errors.  The prototype for virtually all of
the current and proposed 
IACTs is the Whipple Observatory 10~m telescope located
on Mt. Hopkins in southern Arizona \cite{cawley90}.
Newer instruments include CAT, a telescope in
the Pyren\'ees operated by a French collaboration \cite{barrau98},
 CANGAROO, a Japanese-Australian telescope in Woomera,
Australia \cite{hara93},
the HEGRA five telescope array operated
by an Armenian-German-Spanish collaboration on
La Palma in the Canary islands \cite{daum97}.

The STACEE and CELESTE experiments also detect gamma rays by
detecting Cherenkov light.  Instead of imaging the Cherenkov flashes
with an optical telescope, these detectors employ large
arrays of heliostats
at solar power plants to measure the lateral distribution and to minimize
the energy threshold.  STACEE currently uses a number of heliostats
of 37~m$^2$ area each focused onto a PMT located in a central tower and
read out by a GHz flash analog to digital converter.
The direction of the shower front is determined by the relative
signal delays in the PMTs.
This technique provides a huge 
mirror area and hence a substantial improvement (proportional
to the square root of the mirror area) in the signal to noise
ratio for detecting Cherenkov flashes against the night sky background.
The energy threshold is limited by this signal to noise ratio
as well as by practical limitations in the
trigger electronics, but STACEE and CELESTE are
already achieving thresholds below
of 75~GeV. 
Upgrades (now well underway) should result in reduced energy thresholds
$\sim$30~GeV for both instruments.  The shape of the shower
front will provide some rejection of hadronic background, but
this rejection factor is still to be determined and
probably will not reach that of a single imaging
ACT.  But the low energy threshold and large effective area
hold promise for the detection of pulsars, AGNs 
and other objects.

\subsection{Extensive Air Shower Arrays}

\begin{figure}[htb]
\vspace*{0.0in}
\begin{center}
\end{center}
\vspace*{0.0in}
\includegraphics[height=.1\textheight]{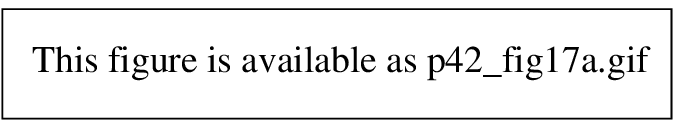}
\includegraphics[height=.1\textheight]{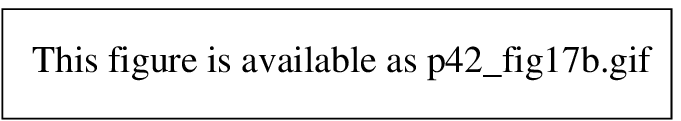}
\caption{Events from two varieties of ground-based
gamma-ray detectors discussed in
the text. {\it Left:} Measured arrival time of particles in an
 extensive air shower
array in the Milagro detector.  The height of the vertical lines
represents the time in nanoseconds.  The lines are to scale
(1 foot = 1 ns). {\it Right:} Events from the 109 pixel camera on the
Whipple telescope (circa 1995).  From left to right, top to bottom the
first image is that of a candidate gamma-ray (pointing to the putative
source the center of the field), an accidental trigger from Poisson
fluctuations in the night sky background
light, a hadronic shower from a
cosmic-ray event, and a ring image from a single muon.}
\label{fig:events}
\end{figure}

Extensive air shower arrays directly detect the particles that reach
the ground.  The direction of the primary gamma ray is determined by
measuring the relative arrival time of the air shower
over a large area.  Figure \ref{fig:events} shows the relative
arrival time of the shower front in the Milagro detector.
Until recently EAS arrays consisted of sparse arrays of plastic
scintillators spread over large areas.  Typical physical coverage was
between 0.5\% (the CYGNUS array) and 1\% (the CASA array).
With such a small sampling of the air shower, thousands of particles
needed to reach the ground to reconstruct the air shower.  These arrays 
where sensitive to primary photons with energies above $\sim100$ TeV.
There are two techniques to lower the energy threshold of EAS arrays.
The technique developed by the Milagro collaboration is to
make the entire detector area active.  This is accomplished at a
reasonable cost by using water as the
detecting medium.  Since the Cherenkov angle in water is $41^{\circ}$
a layer of photomultiplier tubes (PMTs) at a depth comparable to their
spacing can detect a particle that enters anywhere in the detector.
In addition if the layer of water above the PMTs is several radiation
lengths, the more numerous photons in the air shower
will convert to electrons and/or positrons above the PMTs and can be 
detected.  The other approach is to place the array at very high altitude.
As can be seen in Figure \ref{fig:approxb}, nearly five times as many
electrons are present for the same energy shower at 4km asl as at 2.6km asl. 
The Tibet array has pioneered this approach.

\subsubsection{Background Rejection in Extensive Air Shower Arrays}

A fundamental problem in performing gamma-ray astronomy from the ground
is differentiating extensive air showers generated by primary gamma rays
from those generated by hadronic cosmic rays.  The techniques developed
by Weekes and Turver for air Cherenkov telescopes has had great success.
The situation in EAS arrays is not as well developed.  The primary technique
used in EAS arrays has been to detect the penetrating component
(mainly muons and hadrons) present in hadronic air showers.  The Milagro
detector has a second layer of PMTs located beneath 6 meters of water for
this purpose.  The techniques developed to date yield a factor of $\sim1.8$
improvement in sensitivity \cite{sinnis01}, denoted as the $Q$-factor. 
This technique depends on the ability to measure the energy deposited in
the detector and thus favors thick, calorimetric instruments.  
 
\begin{figure}[htb]
  \label{fig:expsense}
  \begin{center}
    \includegraphics[height=.1\textheight]{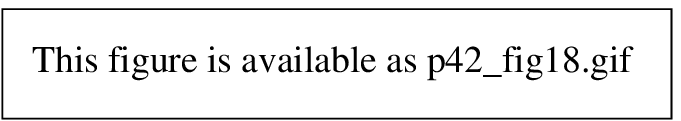}
  \end{center}
  \caption{Expected sensitivity of current and planned instruments.}
\end{figure}

By densely sampling the particles that survive to ground level,
non-calorimetric instruments may also reject the hadronic background.
Gamma-ray induced air showers tend to have a smooth particle distribution with 
a well localized core.  Hadronic showers exhibit clumps of particles on
large and small scales.  Applied to a relatively small detector
(5,000 m$^2$) this technique could give sensitivity improvements comparable
to that achieved with Milagro
\cite{bussino}, $Q\sim1.8$.  Figure \ref{fig:expsense}
shows the expected sensitivities of current and planned instruments.
For the air shower arrays (Milagro, Tibet, and ARGO) the integration time
is 1 year.  For the air Cherenkov instruments the integration time
is 50 hours on the source (which is a typical amount of time spent on a
given source in a year).  The sensitivity shown for ARGO assumes no background rejection,
and should be viewed as a conservative preliminary estimate \cite{Piazzoli}.
 
\section{Future Experimental Directions}

What is the logical follow-up to the current experiments that will
facilitate the best scientific return?  A reduction in
energy threshold of conventional ACTs would increase overlap with
GLAST, and would extend the gamma-ray horizon allowing the detection of
more distant extragalactic sources and GRBs.
However, conventional imaging ACTs
are narrow field instruments for which observations must be guided by
some other source of information (GLAST source locations, X-ray surveys).
For flaring sources, a trigger from a wide-field instrument is required
and the time to slew a large telescope may well exceed the most interesting
timescale.  To achieve sensitivity to the shortest AGN flares, to provide
continuous coverage of the rapidly varying emission, to provide
spectral cutoffs for a large ensemble of AGN, and to catch short and
long-duration GRBs during their development requires a global
network of very wide field instruments
instrument with very good sensitivity to $E>100$ GeV events, and optimized
for temporal coverage rather than point source sensitivity.

The current generation of space-based gamma-ray detectors (AGILE and GLAST)
will provide all-sky coverage and good sensitivity to both point sources and
diffuse emission at energies above 100~MeV.
Above 50 GeV, HESS, VERITAS and other ground-based atmospheric
Cherenkov detectors will provide the high
point source sensitivity and huge effective areas (more than $\sim 10^4$ times
that of satellite experiments) required to study the shortest
variability time-scales and to provide time-resolved
spectra from 50 GeV up to 50 TeV.  Arrays of imaging atmospheric Cherenkov
detectors now under construction (e.g., VERITAS in the northern hemisphere and
HESS in the southern hemisphere)
complement GLAST.  Jointly these experiments are assured to make major
strides in astrophysics.

\subsection{Atmospheric Cherenkov Instruments}

\subsubsection{High Sensitivity: VERITAS and HESS}

VERITAS is a proposed array of seven 10m aperture optical
reflectors that combines the best elements of the large aperture
Whipple telescope, with the power of stereoscopic imaging pioneered
by the HEGRA collaboration.
VERITAS will have good sensitivity in the energy range 50\,GeV to 50\,TeV
energy range.  To quote a typical figure of merit, for observations of
the Crab nebula the peak counting rate will be obtained at about 75 GeV.
Each VERITAS telescopes will use the imaging
concept developed by the Whipple Observatory Gamma-Ray Collaboration
and used to detect the first galactic and extragalactic sources of TeV
$\gamma$-rays but will achieve an order of magnitude better
sensitivity and a significant reduction in energy threshold to 75\,GeV.
The new telescopes will improve upon the design of the existing 10\,m
$\gamma$-ray telescope.  Each telescope will have a tessellated 10\,m
mirror and long focal length (12\,m) with the optical design optimized
for good angular resolution, a large field of view, minimum wavefront
distortion and identical, spherical mirror facets.  Each telescope
camera will have 499 pixels covering a 3.5$^\circ$ field of view with
0.15$^\circ$ resolution.  The detectors and camera electronics will
use high speed 500 MHz flash ADCs, and a low time jitter intelligent
trigger.  The array will
have an effective collection area in excess of 40,000\,m$^2$.  VERITAS
will be located in southern Arizona where it will take advantage of
the existing infrastructure of the Whipple Observatory/  With
stereoscopic imaging, the array will achieve an unprecedented angular
resolution, energy resolution and background rejection over three
decades of energy.  The primary scientific objectives of VERITAS will
be the study of active galactic nuclei, supernova remnants, pulsars,
$\gamma$-ray bursts, and the search for new sources.  The minimum
detectable flux sensitivity will be 0.5\% of the Crab Nebula at 200
GeV, a factor of 20 improvement over the Whipple telescope, the most
sensitive telescope currently operating in this energy range.  The
angular resolution ($<$0.05$^\circ$) will be sufficient to identify a
number of the unidentified sources detected by EGRET on the Compton
Gamma Ray Observatory. A simple extrapolation suggests that more than
30 active galactic nuclei will be detected. VERITAS provides a unique
combination of large collection area, low background, and good energy
resolution. 

\subsubsection{Low Energy Threshold: 5@5}

The ``5@5" Cherenkov telescope array is a design concept for a
ground-based gamma-ray telescope system that will bring the
enormous $\sim 10^4\ {\rm m}^2$ meter collection area that Cherenkov telescopes
enjoy at TeV energies down to the GeV energy range. Currently, the
GeV energy range is accessible only to space-based instruments like
EGRET and the future GLAST. While such instruments have wide fields
of view and thus are well-suited for surveys, they also have
inherently small collection areas ($\sim 1\ {\rm m}^2$) and thus
are not well-suited for monitoring sources
that exhibit strong GeV variability on second
to hour timescales, e.g., gamma-ray bursts and blazars.
Being able to follow this rapid variability is
crucial for constraining the emission mechanisms and physics conditions
in these sources.

Referring to Fig.\ref{fig:expsense}, the sensitivity curves for 5@5
(ground-based) and GLAST (space-based) would cross at 10 GeV. Note,
though, that the source integration times are very different
for the two instruments. The integration times used (50 hours vs. 1 year or
$sim 9000$ hours) are chosen as the maximum total times one can easily
acquire on a source during one observing season. (Ground-based instruments
have a much lower duty cycle because they only operate in
dark conditions.) For the detection of a new, steady source, the
two instruments are comparable in capabilities. However,
the ground-based telescope has acquired its photons in a time interval
that is $\sim$ 200 times shorter -- a crucial advantage when one is dealing
with rapidly varying sources. While a 5@5-class instrument and GLAST
may appear to be competitors and perhaps redundant,
they are in fact highly complementary. Cherenkov arrays are narrow
field instruments and the sky density of transient gamma-ray sources
is very low, i.e., a 5@5 instrument requires a
wide-field instrument like GLAST to tell it where to point.
A low threshold experiment such as  5@5 and
GLAST provide two complementary approaches to achieving some of the
most important scientific objectives: GLAST would be ideal for surveys and 5@5
for deep studies of spectral and temporal characteristics of gamma-ray
sources.  In particular, 5@5 would be an excellent
instrument for study of variable/transient gamma-ray phenomena (blazars,
microquasars, and long duration GRBs) if a wide-field trigger is available.
To put the capabilities of such an instrument in perspective, consider
a transient source with the flux of the Vela pulsar;  
such a GeV transient source could be detected by 5@5 for
just 1 second exposure time!
The rapid development and successful operation of a low-threshold
Cherenkov telescope during the lifetime of GLAST would therefore represent a
major observational coup. 

Another route to such improvements in energy
threshold would be to upgrade VERITAS, HESS or MAGIC with higher
quantum efficiency detectors.  Current employed bialkali PMTs have
quantum efficiencies of $\sim$20\%, while semitransparent InGaN
or reflection-mode GaAsP solid state cathodes promise to double
or triple the efficiency.

To lower the energy threshold, the collection efficiency of
Cherenkov photons must be dramatically improved. Barring
some fundamental breakthrough in detector quantum efficiency
there are only two ways to increase the collection:
(i) build a much larger, more expensive telescope mirror, or (ii)
use smaller, cheaper mirrors but go to higher altitudes where
fewer Cherenkov photons suffer atmospheric absorption.
The first approach represents the strategy employed
by the MAGIC collaboration (a single large 17 meter mirror at
at an altitude of 2000 meters),
while the second is the one behind the proposed 5@5 concept (5 12 meter
mirrors at an altitude of 5000 meters operating in a stereoscopic
configuration).  The advantages of operating at high altitude
are several. First, the amount of Cherenkov light per gamma-ray primary is
higher. This allows one to detect a Cherenkov pulse with a smaller
mirror, and thus it becomes feasible to consider the construction
of several of these mirrors. When placed in a suitable array
configuration (in the 5@5 design, four mirrors are placed at the
corners of a square and one telescope is placed at the center),
these telescopes can be jointly triggered and operated stereoscopically.
Even in the $\sim 10-100$ GeV regime where the gamma-ray shower shape becomes
less well-defined, this allows for a much cleaner rejection of cosmic ray
and muon backgrounds than is possible with a single telescope. An additional
advantage of going to higher altitudes is that the cosmic-ray induced
background is intrinsically lower to begin with because cosmic-ray
showers tend to peak at lower altitudes. At energies below $\sim 100$
GeV and at high altitudes, one can therefore operate in a mode where
effectively background comes only from cosmic ray electrons. This background
can be readily measured and removed in the same way one subtracts the
night sky background in optical CCD images. Note also that at very
low energies this electron background can
be minimized by the appropriate choice of site location. The proposed
5@5 site in Chile has a geomagnetic latitude such that the low-energy
electron rigidity cutoff there is 15 GeV, i.e., the background
at lower energies may be strongly suppressed leading to a corresponding
gain in sensitivity. The proposed location of the 5@5 array in the
Atacama desert plateau also overcomes
the one major disadvantage that has hampered previous attempts at
high altitude arrays, namely the combination of bad weather and the
sheer physical difficulty of operating in a high altitude environment.
The Atacama desert is one of the best astronomical sites in the world
(with stable atmospheric conditions and low cloud cover), and
and it already has excellent support infrastructure because of the large,
international millimeter array (ALMA) that is currently being built there.

\subsection{Large Field-of-View Instruments}

The development of a ground-based gamma ray instrument with a large
field-of-view and good sensitivity (comparable to that projected for
VERITAS) should be a high priority.  This is required for the observation
of flaring AGN and perhaps more importantly to observe the high-energy
emission from gamma ray bursts.  At present there are two possible
approaches.  The development of wide field optics is a requirement of
the OWL instrument (a space-based instrument to detect the fluorescence
in the atmosphere from ultra-high-energy cosmic rays).  Similar technology
could be used in ground-based air Cherenkov telescopes.  Alternatively,
one can build a large area, dense sampling, extensive air
shower array at high altitude.  Both techniques require further development.

\subsubsection{Large Field-of-View Air Cherenkov Instruments}

To achieve simultaneous coverage of a large portion of the sky,
instruments with fields of view of 30 to 60 degrees will be
required.   Reflective optics are limited to about 15 degree fields
of view, above which the secondary (or tertiary) optics grow to the point
that the entire primary is obscured!  Technical innovations such as 
refractive optics (UV transparent Fresnel lenses), large area
megapixel detectors, high speed ASICs, and new data handling capabilities
will be required to make such an instrument.
Limitations in the size of Fresnel optics may also necessitate
higher quantum efficiency detectors to achieve a target threshold 
approaching 100~GeV.  

An impetus for this work comes from the proposal to build
a downward looking atmospheric fluorescence detector (OWL or EUSO) to
detect ultra-high energy cosmic rays and neutrinos.  This technologically
bold proposal would require the deployment of 3 to 10m optics in space,
with megapixel cameras covering the same area.  
Future ground-based gamma-ray experiments and d
space-based atmospheric fluorescence detectors
have some remarkable similarities in their technological requirements.
Both require wide-field (10-60 degree) optics, large apertures
(10m), high quantum efficiency UV/blue photon-counting detectors,
cameras with large plate scales and large numbers of pixels (up
to 1 million) and inexpensive ASICs for time-digitization of signals.
Collaborations between the gamma-ray and ultra-high energy cosmic
ray communities are a logical step to develop the enabling technologies.

\subsubsection{Extensive Air Shower Arrays}

The development of high-sensitivity extensive air shower arrays is 
a nascent field.  To date the Milagro instrument is the only instrument
operating at relatively low energy threshold that has
the ability to reject the cosmic-ray background.  ARGO is an instrument
currently under construction in Tibet, that will also have the ability to
reject the cosmic-ray background.  Neither instrument approaches the
rejection levels now attained in air Cherenkov telescopes.  This
represents the major challenge for future air shower arrays.  The goals
of the next generation extensive air shower arrays are to:

\begin{enumerate}

\item reach sensitivities (with one year of integration) comparable to
those achieved by VERITAS (with 50 hours of on-source integration), 

\item have sufficient effective area at or below 100 GeV to see more
distant gamma ray bursts (if their spectra continue to these energies).

\end{enumerate}
These goals require that current instruments must undergo a factor of
$\sim 50-100$ improvement in sensitivity.
The sensitivity to a point source can be calculated with the following
formula.
\begin{equation}
S = Q\sigma_{\theta} \frac{\int_{E_{th}}^{\inf} A_{\gamma}(E) E^{-s}dE} 
             {\int_{E_{th}}^{\inf} A_{p}E^{-2.7}dE}
\end{equation}
where $s$ is the differential spectral index of the source,
$A_{\gamma ,p}(E)$ is the effective area for gamma rays/protons,
$\sigma_{theta}$ is the angular resolution of the instrument, and $Q$
is the gain in sensitivity due to background rejection
($\epsilon_{\gamma}/\sqrt(\epsilon_{p})$,
where $\epsilon_{\gamma ,p}$ is the efficiency for keeping gamma rays/protons).  
For a Crab like source ($s=2.4$) and with the 
assumption that the effective area is a step function one can derive
the following figure of merit for EAS arrays:
\begin{equation}
\label{eqn:merit}
M \approx Q\sigma_{\theta} \frac{R_{\gamma} \sqrt{A_{det}}} {\sqrt{E_{th}}}
\end{equation}
where $R_{\gamma}$ is the square root of the ratio of effective areas
for gamma rays and protons.  The fact that $R_{\gamma}$ is not unity
is a reflection of the differences in the development of gamma ray and
proton induced air showers.  In general $R_{\gamma}$ is a function of
energy and altitude (or zenith angle of observation).  For Milagro,
located at 2600 m asl, $R_{\gamma} \sim 1$.  At 4000m asl $R_{\gamma}$
is between 1.5 and 2.  The dependence on $E_{th}$ given above is
dependent upon the source spectrum.

Examination of Figure \ref{fig:approxb} shows that for a 100 GeV gamma
ray induced air shower there are $\sim$5 times as many electrons and
positrons at 4000m asl than at 2600m asl.  It is clear that any future
detector must be located at extreme altitudes.  The next requirement is
driven by the fact that there are many more photons than electrons present
in the air shower.  This demands that the detector be ``thick", several
radiation lengths, to observe these photons.  Using equation \ref{eqn:merit}
one can determine that a Milagro type detector located at 4000m asl and
with $10\times$ the physical area will be roughly 10-15 times more sensitive
than the current Milagro instrument.  However, the operational
experience of Milagro has highlighted several areas where large improvements
can be made.  These issues are currently being addressed but we believe
that they will lead to a factor of 3 improvement in sensitivity.
The biggest issue is simply the size of the detector.  With an area
5000 m$^2$ and high sensitivity to individual particles, most of the 
showers ($80\%$) that trigger the detector have their cores outside of
the pond.  This leads to a degradation in the angular resolution of the
instrument.  An array of 170 water tanks is currently being built around
the main pond to locate the shower core for these events.  Monte Carlo
simulations indicate that this should improve the sensitivity of Milagro
by a factor of 2, this comes both from improved angular resolution and
improved background rejection.  In addition, the main impediment to
lowering the trigger threshold of Milagro is the rate of
single muons.  Smart triggers should enable one to lower the energy
threshold to further improve the sensitivity by 30\%.
Thus, there is good reason to expect that such an instrument located at
4000m asl will have a sensitivity 
$\sim$30 times greater than the current Milagro instrument. 
To make further improvements one needs to improve the background
rejection and/or the angular resolution of the
instrument.  Given the relative infancy of background rejection in
air shower arrays, and the great progress made
by air Cherenkov instruments in the past decade, this does not seem
to be an unreasonable improvement.  However,
work is needed to demonstrate that such an improvement can be realized.

\section{Conclusions}
The field of gamma-ray astronomy has changed radically in the past decade. In 
1990 there were a handful of gamma-ray sources observed from space and only
one source observed from the ground.  Today there are over 150 known sources
of gamma rays observed from space and about 10 sources of TeV gamma rays
observed from the ground. The coming decade promises even greater advances.
There are new ground-based and space-based instruments
currently under development that will increase the number of known sources
by over an order of magnitude.  Large field-of-view instruments with
relatively low-energy thresholds ($\sim 500$ GeV) capable of detecting
very-high-energy emission from gamma-ray bursts are now beginning to
acquire data.  It is clear that this is still a relatively young field
were new ideas are being pursued and large advances are on the horizon.

At the workshop we heard from many participants about current measurements
and the increasing overlap with particle physics.  The time where distant
cosmic sources can be used to probe
the fundamental interactions of matter at energy scales unattainable on
Earth is approaching.  
Much discussion centered around possible future directions.  Two goals
were put forth for ground-based instruments: a very low energy threshold
instrument ($\sim$10 GeV) and a very sensitive wide-field
instrument with an energy threshold $\sim$100 GeV.  We look forward to
the coming decade.

\section{References}

\bibliography{p42}

\begin{thebibliography}{79}
\expandafter\ifx\csname natexlab\endcsname\relax\def\natexlab#1{#1}\fi
\expandafter\ifx\csname bibnamefont\endcsname\relax
  \def\bibnamefont#1{#1}\fi
\expandafter\ifx\csname bibfnamefont\endcsname\relax
  \def\bibfnamefont#1{#1}\fi
\expandafter\ifx\csname citenamefont\endcsname\relax
  \def\citenamefont#1{#1}\fi
\expandafter\ifx\csname url\endcsname\relax
  \def\url#1{\texttt{#1}}\fi
\expandafter\ifx\csname urlprefix\endcsname\relax\def\urlprefix{URL }\fi
\providecommand{\bibinfo}[2]{#2}
\providecommand{\eprint}[2][]{\url{#2}}

\bibitem[{\citenamefont{Hartman and et~al.}(1999)}]{Hartman99}
\bibinfo{author}{\bibfnamefont{R.~C.} \bibnamefont{Hartman}} \bibnamefont{and}
  \bibinfo{author}{\bibnamefont{et~al.}}, \bibinfo{journal}{ApJS}
  \textbf{\bibinfo{volume}{123}}, \bibinfo{pages}{79} (\bibinfo{year}{1999}).

\bibitem[{\citenamefont{Weekes and Turver}(1977)}]{weekes-img}
\bibinfo{author}{\bibfnamefont{T.~C.} \bibnamefont{Weekes}} \bibnamefont{and}
  \bibinfo{author}{\bibfnamefont{K.~E.} \bibnamefont{Turver}}, in
  \emph{\bibinfo{booktitle}{in {\it Proc. 12th ESLAB Symp.}}}
  (\bibinfo{address}{Frascati, ESA Sp-124}, \bibinfo{year}{1977}), p.
  \bibinfo{pages}{279}.

\bibitem[{\citenamefont{von Montigny and et~al.}(1995)}]{Montigny95}
\bibinfo{author}{\bibfnamefont{C.}~\bibnamefont{von Montigny}}
  \bibnamefont{and} \bibinfo{author}{\bibnamefont{et~al.}},
  \bibinfo{journal}{ApJ} \textbf{\bibinfo{volume}{440}}, \bibinfo{pages}{525}
  (\bibinfo{year}{1995}).

\bibitem[{\citenamefont{Sikora et~al.}(1994)\citenamefont{Sikora, Begelman, and
  Rees}}]{Sikora94}
\bibinfo{author}{\bibfnamefont{M.}~\bibnamefont{Sikora}},
  \bibinfo{author}{\bibfnamefont{M.}~\bibnamefont{Begelman}}, \bibnamefont{and}
  \bibinfo{author}{\bibfnamefont{M.}~\bibnamefont{Rees}},
  \bibinfo{journal}{ApJ} \textbf{\bibinfo{volume}{421}}, \bibinfo{pages}{153}
  (\bibinfo{year}{1994}).

\bibitem[{\citenamefont{Mannheim}(1993)}]{mannheim93}
\bibinfo{author}{\bibfnamefont{K.}~\bibnamefont{Mannheim}},
  \bibinfo{journal}{A\&A} \textbf{\bibinfo{volume}{269}}, \bibinfo{pages}{67}
  (\bibinfo{year}{1993}).

\bibitem[{\citenamefont{Kerrick and et~al.}(1995b)}]{Kerrick95b}
\bibinfo{author}{\bibfnamefont{A.}~\bibnamefont{Kerrick}} \bibnamefont{and}
  \bibinfo{author}{\bibnamefont{et~al.}}, \bibinfo{journal}{ApJ}
  \textbf{\bibinfo{volume}{452}}, \bibinfo{pages}{588} (\bibinfo{year}{1995b}).

\bibitem[{\citenamefont{Quinn and et~al.}(1995)}]{Quinn95}
\bibinfo{author}{\bibfnamefont{J.}~\bibnamefont{Quinn}} \bibnamefont{and}
  \bibinfo{author}{\bibnamefont{et~al.}}, in \emph{\bibinfo{booktitle}{in {\it
  Proc. 24th ICRC}}} (\bibinfo{address}{Rome}, \bibinfo{year}{1995}),
  vol.~\bibinfo{volume}{2}, p. \bibinfo{pages}{369}.

\bibitem[{\citenamefont{Petry and et~al.}(1997)}]{Petry97}
\bibinfo{author}{\bibfnamefont{D.}~\bibnamefont{Petry}} \bibnamefont{and}
  \bibinfo{author}{\bibnamefont{et~al.}}, in \emph{\bibinfo{booktitle}{in {\it
  Proc. 25th ICRC}}} (\bibinfo{address}{Durban, South Africa},
  \bibinfo{year}{1997}), vol.~\bibinfo{volume}{3}, p. \bibinfo{pages}{241}.

\bibitem[{\citenamefont{Punch and et~al.}(1992)}]{Punch92}
\bibinfo{author}{\bibfnamefont{M.}~\bibnamefont{Punch}} \bibnamefont{and}
  \bibinfo{author}{\bibnamefont{et~al.}}, \bibinfo{journal}{Nature}
  \textbf{\bibinfo{volume}{358}}, \bibinfo{pages}{477} (\bibinfo{year}{1992}).

\bibitem[{\citenamefont{Quinn and et~al.}(1996)}]{Quinn96}
\bibinfo{author}{\bibfnamefont{J.}~\bibnamefont{Quinn}} \bibnamefont{and}
  \bibinfo{author}{\bibnamefont{et~al.}}, \bibinfo{journal}{ApJ}
  \textbf{\bibinfo{volume}{456}}, \bibinfo{pages}{L83} (\bibinfo{year}{1996}).

\bibitem[{\citenamefont{Catanese}(1998)}]{Catanese98}
\bibinfo{author}{\bibfnamefont{M.}~\bibnamefont{Catanese}},
  \bibinfo{journal}{ApJ} \textbf{\bibinfo{volume}{501}}, \bibinfo{pages}{616}
  (\bibinfo{year}{1998}).

\bibitem[{\citenamefont{Chadwick and et~al.}(1999)}]{chadwick99}
\bibinfo{author}{\bibfnamefont{P.~M.} \bibnamefont{Chadwick}} \bibnamefont{and}
  \bibinfo{author}{\bibnamefont{et~al.}}, \bibinfo{journal}{Astropart. Phys.}
  \textbf{\bibinfo{volume}{11}}, \bibinfo{pages}{145} (\bibinfo{year}{1999}).

\bibitem[{\citenamefont{Nishiyama and et~al.}(1999)}]{Nishiyama99}
\bibinfo{author}{\bibnamefont{Nishiyama}} \bibnamefont{and}
  \bibinfo{author}{\bibnamefont{et~al.}}, in \emph{\bibinfo{booktitle}{in {\it
  Proc. 26th ICRC}}} (\bibinfo{address}{Salt Lake City, Utah},
  \bibinfo{year}{1999}), \bibinfo{note}{(unpublished communication)}.

\bibitem[{\citenamefont{Horan and et~al.}(2001)}]{Horan01}
\bibinfo{author}{\bibfnamefont{D.}~\bibnamefont{Horan}} \bibnamefont{and}
  \bibinfo{author}{\bibnamefont{et~al.}}, in \emph{\bibinfo{booktitle}{in {\it
  Proc. 27th ICRC}}} (\bibinfo{address}{Hamburg, Germany},
  \bibinfo{year}{2001}), p. \bibinfo{pages}{(in press)}.

\bibitem[{\citenamefont{Padovani and Giommi}(1995)}]{padovani95}
\bibinfo{author}{\bibfnamefont{P.}~\bibnamefont{Padovani}} \bibnamefont{and}
  \bibinfo{author}{\bibfnamefont{P.}~\bibnamefont{Giommi}},
  \bibinfo{journal}{ApJ} \textbf{\bibinfo{volume}{444}}, \bibinfo{pages}{567}
  (\bibinfo{year}{1995}).

\bibitem[{\citenamefont{Dermer et~al.}(1992)\citenamefont{Dermer, Schlickeiser,
  and Mastichiadis}}]{Dermer92b}
\bibinfo{author}{\bibfnamefont{C.~D.} \bibnamefont{Dermer}},
  \bibinfo{author}{\bibfnamefont{R.}~\bibnamefont{Schlickeiser}},
  \bibnamefont{and}
  \bibinfo{author}{\bibfnamefont{A.}~\bibnamefont{Mastichiadis}},
  \bibinfo{journal}{A\&A} \textbf{\bibinfo{volume}{256}}, \bibinfo{pages}{L27}
  (\bibinfo{year}{1992}).

\bibitem[{\citenamefont{Gould and Schreder}(1967)}]{Gould67}
\bibinfo{author}{\bibfnamefont{R.}~\bibnamefont{Gould}} \bibnamefont{and}
  \bibinfo{author}{\bibfnamefont{G.}~\bibnamefont{Schreder}},
  \bibinfo{journal}{Phys. Rev.} \textbf{\bibinfo{volume}{155}},
  \bibinfo{pages}{1408} (\bibinfo{year}{1967}).

\bibitem[{\citenamefont{Stecker and Jager}(1993)}]{Stecker93}
\bibinfo{author}{\bibfnamefont{F.}~\bibnamefont{Stecker}} \bibnamefont{and}
  \bibinfo{author}{\bibfnamefont{O.~D.} \bibnamefont{Jager}},
  \bibinfo{journal}{ApJ} \textbf{\bibinfo{volume}{415}}, \bibinfo{pages}{L71}
  (\bibinfo{year}{1993}).

\bibitem[{\citenamefont{Fossati et~al.}(1998)\citenamefont{Fossati, Maraschi,
  Celotti, Comastri, and Ghisellini}}]{fossati98}
\bibinfo{author}{\bibfnamefont{G.}~\bibnamefont{Fossati}},
  \bibinfo{author}{\bibfnamefont{L.}~\bibnamefont{Maraschi}},
  \bibinfo{author}{\bibfnamefont{A.}~\bibnamefont{Celotti}},
  \bibinfo{author}{\bibfnamefont{A.}~\bibnamefont{Comastri}}, \bibnamefont{and}
  \bibinfo{author}{\bibfnamefont{G.}~\bibnamefont{Ghisellini}},
  \bibinfo{journal}{MNRAS} \textbf{\bibinfo{volume}{299}}, \bibinfo{pages}{433}
  (\bibinfo{year}{1998}).

\bibitem[{\citenamefont{Buckley}(2001)}]{Buckley01}
\bibinfo{author}{\bibfnamefont{J.}~\bibnamefont{Buckley}}, in
  \emph{\bibinfo{booktitle}{in {\it Proc. Gamma-2001 Conference}}}
  (\bibinfo{publisher}{AIP}, \bibinfo{address}{Baltimore, MD},
  \bibinfo{year}{2001}), p. \bibinfo{pages}{(in press)}.

\bibitem[{\citenamefont{Buckley and et~al.}(1996)}]{Buckley96}
\bibinfo{author}{\bibfnamefont{J.}~\bibnamefont{Buckley}} \bibnamefont{and}
  \bibinfo{author}{\bibnamefont{et~al.}}, \bibinfo{journal}{ApJ}
  \textbf{\bibinfo{volume}{472}}, \bibinfo{pages}{L9} (\bibinfo{year}{1996}).

\bibitem[{\citenamefont{Jordan and et~al.}(2001)}]{Jordan01}
\bibinfo{author}{\bibfnamefont{M.}~\bibnamefont{Jordan}} \bibnamefont{and}
  \bibinfo{author}{\bibnamefont{et~al.}}, in \emph{\bibinfo{booktitle}{in {\it
  Proc. 27th ICRC}}} (\bibinfo{address}{Hamburg, Germany},
  \bibinfo{year}{2001}), p. \bibinfo{pages}{(in press)}.

\bibitem[{\citenamefont{Fossati and et~al.}(2001)}]{Fossati01}
\bibinfo{author}{\bibfnamefont{G.}~\bibnamefont{Fossati}} \bibnamefont{and}
  \bibinfo{author}{\bibnamefont{et~al.}} (\bibinfo{year}{2001}),
  \bibinfo{note}{(in preparation)}.

\bibitem[{\citenamefont{Krawczynski et~al.}(2000)\citenamefont{Krawczynski,
  Coppi, Maccarone, and Aharonian}}]{Krawczynski00}
\bibinfo{author}{\bibfnamefont{H.}~\bibnamefont{Krawczynski}},
  \bibinfo{author}{\bibfnamefont{P.~S.} \bibnamefont{Coppi}},
  \bibinfo{author}{\bibfnamefont{T.}~\bibnamefont{Maccarone}},
  \bibnamefont{and} \bibinfo{author}{\bibfnamefont{F.~A.}
  \bibnamefont{Aharonian}}, \bibinfo{journal}{A\& A}
  \textbf{\bibinfo{volume}{353}}, \bibinfo{pages}{97} (\bibinfo{year}{2000}).

\bibitem[{\citenamefont{Bahcall and Waxman}(2001)}]{waxman-bahcall-01}
\bibinfo{author}{\bibfnamefont{J.}~\bibnamefont{Bahcall}} \bibnamefont{and}
  \bibinfo{author}{\bibfnamefont{E.}~\bibnamefont{Waxman}},
  \bibinfo{journal}{Phys Rev D} \textbf{\bibinfo{volume}{61}},
  \bibinfo{pages}{3002} (\bibinfo{year}{2001}).

\bibitem[{\citenamefont{Aharonian}(2000)}]{Aharonian00}
\bibinfo{author}{\bibfnamefont{F.}~\bibnamefont{Aharonian}},
  \bibinfo{journal}{New Astronomy} \textbf{\bibinfo{volume}{5}},
  \bibinfo{pages}{377} (\bibinfo{year}{2000}).

\bibitem[{\citenamefont{Catanese and et~al.}(1997a)}]{Catanese97a}
\bibinfo{author}{\bibfnamefont{M.}~\bibnamefont{Catanese}} \bibnamefont{and}
  \bibinfo{author}{\bibnamefont{et~al.}}, \bibinfo{journal}{ApJ}
  \textbf{\bibinfo{volume}{487}}, \bibinfo{pages}{L143}
  (\bibinfo{year}{1997a}).

\bibitem[{\citenamefont{Buckley and et~al.}(1997)}]{Buckley97}
\bibinfo{author}{\bibfnamefont{J.}~\bibnamefont{Buckley}} \bibnamefont{and}
  \bibinfo{author}{\bibnamefont{et~al.}}, in \emph{\bibinfo{booktitle}{in {\it
  Proc. of the 4th Compton Symp.}}}, edited by
  \bibinfo{editor}{\bibfnamefont{C.}~\bibnamefont{Dermer}},
  \bibinfo{editor}{\bibfnamefont{M.}~\bibnamefont{Strickman}},
  \bibnamefont{and} \bibinfo{editor}{\bibfnamefont{J.}~\bibnamefont{Kurfess}}
  (\bibinfo{publisher}{AIP}, \bibinfo{year}{1997}), vol. \bibinfo{volume}{410},
  p. \bibinfo{pages}{1381}.

\bibitem[{\citenamefont{Klebesadel et~al.}(1973)\citenamefont{Klebesadel,
  Strong, and Olson}}]{klebs}
\bibinfo{author}{\bibfnamefont{R.~W.} \bibnamefont{Klebesadel}},
  \bibinfo{author}{\bibfnamefont{I.~B.} \bibnamefont{Strong}},
  \bibnamefont{and} \bibinfo{author}{\bibfnamefont{R.~A.} \bibnamefont{Olson}},
  \bibinfo{journal}{Astrophys. J} \textbf{\bibinfo{volume}{182}},
  \bibinfo{pages}{L85} (\bibinfo{year}{1973}).

\bibitem[{\citenamefont{Costa and et~al.}(1997)}]{costa}
\bibinfo{author}{\bibfnamefont{E.}~\bibnamefont{Costa}} \bibnamefont{and}
  \bibinfo{author}{\bibnamefont{et~al.}}, \bibinfo{journal}{Nature}
  \textbf{\bibinfo{volume}{387}}, \bibinfo{pages}{783} (\bibinfo{year}{1997}).

\bibitem[{\citenamefont{Frail and et~al.}(2001)}]{frail}
\bibinfo{author}{\bibfnamefont{D.~A.} \bibnamefont{Frail}} \bibnamefont{and}
  \bibinfo{author}{\bibnamefont{et~al.}} (\bibinfo{year}{2001}),
  \eprint{astro-ph/0102282}.

\bibitem[{\citenamefont{Norris et~al.}(2000)\citenamefont{Norris, Marani, and
  Bonnell}}]{norris99}
\bibinfo{author}{\bibfnamefont{J.~P.} \bibnamefont{Norris}},
  \bibinfo{author}{\bibfnamefont{G.}~\bibnamefont{Marani}}, \bibnamefont{and}
  \bibinfo{author}{\bibfnamefont{J.~T.} \bibnamefont{Bonnell}},
  \bibinfo{journal}{Astrophys. J.} \textbf{\bibinfo{volume}{534}},
  \bibinfo{pages}{248} (\bibinfo{year}{2000}).

\bibitem[{\citenamefont{Hurley and et~al.}(1994)}]{hurley}
\bibinfo{author}{\bibfnamefont{K.}~\bibnamefont{Hurley}} \bibnamefont{and}
  \bibinfo{author}{\bibnamefont{et~al.}}, \bibinfo{journal}{Nature}
  \textbf{\bibinfo{volume}{372}}, \bibinfo{pages}{652} (\bibinfo{year}{1994}).

\bibitem[{\citenamefont{Jungman et~al.}(1995)\citenamefont{Jungman,
  Kamionkowski, and Greist}}]{jung-kamio-greist-95}
\bibinfo{author}{\bibfnamefont{G.}~\bibnamefont{Jungman}},
  \bibinfo{author}{\bibfnamefont{M.}~\bibnamefont{Kamionkowski}},
  \bibnamefont{and} \bibinfo{author}{\bibfnamefont{K.}~\bibnamefont{Greist}},
  \bibinfo{journal}{Phys. Reports} \textbf{\bibinfo{volume}{267}},
  \bibinfo{pages}{195} (\bibinfo{year}{1995}).

\bibitem[{\citenamefont{Ellis and et~al.}(1984)}]{ellis-84}
\bibinfo{author}{\bibfnamefont{J.}~\bibnamefont{Ellis}} \bibnamefont{and}
  \bibinfo{author}{\bibnamefont{et~al.}}, \bibinfo{journal}{Nucl. Phys.}
  \textbf{\bibinfo{volume}{B238}}, \bibinfo{pages}{453} (\bibinfo{year}{1984}).

\bibitem[{\citenamefont{Bergstroem et~al.}(1999)\citenamefont{Bergstroem,
  Edsjo, and Ullio}}]{berg-edsjo-ullio-99}
\bibinfo{author}{\bibfnamefont{L.}~\bibnamefont{Bergstroem}},
  \bibinfo{author}{\bibfnamefont{J.}~\bibnamefont{Edsjo}}, \bibnamefont{and}
  \bibinfo{author}{\bibfnamefont{P.}~\bibnamefont{Ullio}},
  \bibinfo{journal}{ApJ} \textbf{\bibinfo{volume}{526}}, \bibinfo{pages}{215}
  (\bibinfo{year}{1999}).

\bibitem[{\citenamefont{Barwick and et~al.}(1997)}]{barwick-97}
\bibinfo{author}{\bibfnamefont{S.}~\bibnamefont{Barwick}} \bibnamefont{and}
  \bibinfo{author}{\bibnamefont{et~al.}}, \bibinfo{journal}{ApJ}
  \textbf{\bibinfo{volume}{482}}, \bibinfo{pages}{L191} (\bibinfo{year}{1997}).

\bibitem[{\citenamefont{Coutu and et~al.}(2001)}]{coutu-01}
\bibinfo{author}{\bibfnamefont{S.}~\bibnamefont{Coutu}} \bibnamefont{and}
  \bibinfo{author}{\bibnamefont{et~al.}}, in \emph{\bibinfo{booktitle}{in {\it
  Proc. 27th ICRC}}} (\bibinfo{address}{Hamburg, Germany},
  \bibinfo{year}{2001}), p. \bibinfo{pages}{(in press)}.

\bibitem[{\citenamefont{Baltz and et~al.}(2001)}]{baltz-01}
\bibinfo{author}{\bibfnamefont{E.}~\bibnamefont{Baltz}} \bibnamefont{and}
  \bibinfo{author}{\bibnamefont{et~al.}} (\bibinfo{year}{2001}),
  \eprint{astro-ph/0109318}.

\bibitem[{\citenamefont{Bergstrom et~al.}(1998)\citenamefont{Bergstrom, Ullio,
  and Buckley}}]{berg-ulli-buck-98}
\bibinfo{author}{\bibfnamefont{L.}~\bibnamefont{Bergstrom}},
  \bibinfo{author}{\bibfnamefont{P.}~\bibnamefont{Ullio}}, \bibnamefont{and}
  \bibinfo{author}{\bibfnamefont{J.}~\bibnamefont{Buckley}},
  \bibinfo{journal}{Astropart. Phys.} \textbf{\bibinfo{volume}{9}},
  \bibinfo{pages}{137} (\bibinfo{year}{1998}).

\bibitem[{\citenamefont{C{\^o}t{\'e} et~al.}(2000)\citenamefont{C{\^o}t{\'e},
  Carignan, and Freeman}}]{cote-etal-00}
\bibinfo{author}{\bibfnamefont{P.}~\bibnamefont{C{\^o}t{\'e}}},
  \bibinfo{author}{\bibfnamefont{C.}~\bibnamefont{Carignan}}, \bibnamefont{and}
  \bibinfo{author}{\bibfnamefont{K.~C.} \bibnamefont{Freeman}},
  \bibinfo{journal}{AJ} \textbf{\bibinfo{volume}{120}}, \bibinfo{pages}{3027}
  (\bibinfo{year}{2000}).

\bibitem[{\citenamefont{de~Blok et~al.}(2001)\citenamefont{de~Blok, McGaugh,
  Bosma, and Rubin}}]{debl-etal-01}
\bibinfo{author}{\bibfnamefont{W.~J.~G.} \bibnamefont{de~Blok}},
  \bibinfo{author}{\bibfnamefont{S.~S.} \bibnamefont{McGaugh}},
  \bibinfo{author}{\bibfnamefont{A.}~\bibnamefont{Bosma}}, \bibnamefont{and}
  \bibinfo{author}{\bibfnamefont{V.~C.} \bibnamefont{Rubin}},
  \bibinfo{journal}{Ap. J.} \textbf{\bibinfo{volume}{552}},
  \bibinfo{pages}{L23} (\bibinfo{year}{2001}).

\bibitem[{\citenamefont{Blais-Ouellette
  et~al.}(2000)\citenamefont{Blais-Ouellette, Amram, and
  Carignan}}]{blai-etal-01}
\bibinfo{author}{\bibfnamefont{S.}~\bibnamefont{Blais-Ouellette}},
  \bibinfo{author}{\bibfnamefont{P.}~\bibnamefont{Amram}}, \bibnamefont{and}
  \bibinfo{author}{\bibfnamefont{C.}~\bibnamefont{Carignan}},
  \bibinfo{journal}{AJ} \textbf{\bibinfo{volume}{121}}, \bibinfo{pages}{1952}
  (\bibinfo{year}{2000}).

\bibitem[{\citenamefont{van~den Bosch and Swaters}(2001)}]{vdbo-swat-01}
\bibinfo{author}{\bibfnamefont{F.~C.} \bibnamefont{van~den Bosch}}
  \bibnamefont{and} \bibinfo{author}{\bibfnamefont{R.~A.}
  \bibnamefont{Swaters}}, \bibinfo{journal}{MNRAS} \textbf{\bibinfo{volume}{\bf
  325}}, \bibinfo{pages}{1017} (\bibinfo{year}{2001}).

\bibitem[{\citenamefont{Font and Navarro}(2001)}]{font-nava-01}
\bibinfo{author}{\bibfnamefont{A.~S.} \bibnamefont{Font}} \bibnamefont{and}
  \bibinfo{author}{\bibfnamefont{J.~F.} \bibnamefont{Navarro}}
  (\bibinfo{year}{2001}), \eprint{astro-ph/0106268}.

\bibitem[{\citenamefont{Ullio et~al.}(2001)\citenamefont{Ullio, Zhao, and
  Kamionkowski}}]{ulli-zhao-kamio-01}
\bibinfo{author}{\bibfnamefont{P.}~\bibnamefont{Ullio}},
  \bibinfo{author}{\bibfnamefont{H.~S.} \bibnamefont{Zhao}}, \bibnamefont{and}
  \bibinfo{author}{\bibfnamefont{M.}~\bibnamefont{Kamionkowski}},
  \bibinfo{journal}{Phys. Rev. D.} \textbf{\bibinfo{volume}{64}},
  \bibinfo{pages}{043504} (\bibinfo{year}{2001}).

\bibitem[{\citenamefont{Gondolo}(2000)}]{gond-00}
\bibinfo{author}{\bibfnamefont{P.}~\bibnamefont{Gondolo}},
  \bibinfo{journal}{Phys. Lett.} \textbf{\bibinfo{volume}{B494}},
  \bibinfo{pages}{181} (\bibinfo{year}{2000}).

\bibitem[{\citenamefont{Gondolo and Silk}(1999)}]{gond-silk-99}
\bibinfo{author}{\bibfnamefont{P.}~\bibnamefont{Gondolo}} \bibnamefont{and}
  \bibinfo{author}{\bibfnamefont{J.}~\bibnamefont{Silk}},
  \bibinfo{journal}{Phys. Rev. Lett.} \textbf{\bibinfo{volume}{83}},
  \bibinfo{pages}{1719} (\bibinfo{year}{1999}).

\bibitem[{\citenamefont{Bahcall and Soneira}(1980)}]{bahc-sone-80}
\bibinfo{author}{\bibfnamefont{J.~N.} \bibnamefont{Bahcall}} \bibnamefont{and}
  \bibinfo{author}{\bibfnamefont{R.~M.} \bibnamefont{Soneira}},
  \bibinfo{journal}{Ap. J. Suppl.} \textbf{\bibinfo{volume}{44}},
  \bibinfo{pages}{73} (\bibinfo{year}{1980}).

\bibitem[{\citenamefont{Persic et~al.}(1996)\citenamefont{Persic, Salucci, and
  Stel}}]{pers-salu-96}
\bibinfo{author}{\bibfnamefont{M.}~\bibnamefont{Persic}},
  \bibinfo{author}{\bibfnamefont{P.}~\bibnamefont{Salucci}}, \bibnamefont{and}
  \bibinfo{author}{\bibfnamefont{F.}~\bibnamefont{Stel}},
  \bibinfo{journal}{MNRAS} \textbf{\bibinfo{volume}{281}}, \bibinfo{pages}{27}
  (\bibinfo{year}{1996}).

\bibitem[{\citenamefont{Navarro et~al.}(1996)\citenamefont{Navarro, Frenk, and
  White}}]{nava-fren-96}
\bibinfo{author}{\bibfnamefont{J.~F.} \bibnamefont{Navarro}},
  \bibinfo{author}{\bibfnamefont{C.}~\bibnamefont{Frenk}}, \bibnamefont{and}
  \bibinfo{author}{\bibfnamefont{S.}~\bibnamefont{White}},
  \bibinfo{journal}{Ap. J.} \textbf{\bibinfo{volume}{462}},
  \bibinfo{pages}{563} (\bibinfo{year}{1996}).

\bibitem[{\citenamefont{Moore et~al.}(1998)}]{moor-etal-98}
\bibinfo{author}{\bibfnamefont{B.}~\bibnamefont{Moore}} \bibnamefont{et~al.},
  \bibinfo{journal}{Ap. J.} \textbf{\bibinfo{volume}{499}}, \bibinfo{pages}{L5}
  (\bibinfo{year}{1998}).

\bibitem[{\citenamefont{Kamionkowski and Liddle}(2000)}]{kami-lidd-00}
\bibinfo{author}{\bibfnamefont{M.}~\bibnamefont{Kamionkowski}}
  \bibnamefont{and} \bibinfo{author}{\bibfnamefont{A.}~\bibnamefont{Liddle}},
  \bibinfo{journal}{Phys. Rev. Lett.} \textbf{\bibinfo{volume}{84}},
  \bibinfo{pages}{4525} (\bibinfo{year}{2000}).

\bibitem[{\citenamefont{Binney et~al.}(2001)\citenamefont{Binney, Gerhardt, and
  Silk}}]{binn-etal-01}
\bibinfo{author}{\bibfnamefont{J.}~\bibnamefont{Binney}},
  \bibinfo{author}{\bibfnamefont{O.}~\bibnamefont{Gerhardt}}, \bibnamefont{and}
  \bibinfo{author}{\bibfnamefont{J.}~\bibnamefont{Silk}},
  \bibinfo{journal}{MNRAS} \textbf{\bibinfo{volume}{321}}, \bibinfo{pages}{471}
  (\bibinfo{year}{2001}).

\bibitem[{\citenamefont{Weinberg and Katz}(2001)}]{wein-katz-01}
\bibinfo{author}{\bibfnamefont{M.~D.} \bibnamefont{Weinberg}} \bibnamefont{and}
  \bibinfo{author}{\bibfnamefont{N.}~\bibnamefont{Katz}}
  (\bibinfo{year}{2001}), \eprint{astro-ph/0110632}.

\bibitem[{\citenamefont{Milosavljevic et~al.}(2001)\citenamefont{Milosavljevic,
  Merritt, Rest, and van~den Bosch}}]{milo-etal-01}
\bibinfo{author}{\bibfnamefont{M.}~\bibnamefont{Milosavljevic}},
  \bibinfo{author}{\bibfnamefont{D.}~\bibnamefont{Merritt}},
  \bibinfo{author}{\bibfnamefont{A.}~\bibnamefont{Rest}}, \bibnamefont{and}
  \bibinfo{author}{\bibfnamefont{F.~C.} \bibnamefont{van~den Bosch}}
  (\bibinfo{year}{2001}), \eprint{astro-ph/0110185}.

\bibitem[{\citenamefont{Alexander}(1999)}]{alex-99}
\bibinfo{author}{\bibfnamefont{T.}~\bibnamefont{Alexander}},
  \bibinfo{journal}{Ap. J.} \textbf{\bibinfo{volume}{527}},
  \bibinfo{pages}{835} (\bibinfo{year}{1999}).

\bibitem[{\citenamefont{Kosack and et~al.}(2001)}]{kosack-01}
\bibinfo{author}{\bibfnamefont{K.}~\bibnamefont{Kosack}} \bibnamefont{and}
  \bibinfo{author}{\bibnamefont{et~al.}}, in \emph{\bibinfo{booktitle}{in {\it
  Proc. 27th ICRC}}}, edited by
  \bibinfo{editor}{\bibfnamefont{M.}~\bibnamefont{Simon}},
  \bibinfo{editor}{\bibfnamefont{E.}~\bibnamefont{Lorenz}}, \bibnamefont{and}
  \bibinfo{editor}{\bibfnamefont{M.}~\bibnamefont{Pohl}}
  (\bibinfo{address}{Hamburg, Germany}, \bibinfo{year}{2001}), p.
  \bibinfo{pages}{2989}.

\bibitem[{\citenamefont{Mayer-Hasselwander and
  et~al.}(1998)}]{mayer-hasselwander-98}
\bibinfo{author}{\bibfnamefont{H.}~\bibnamefont{Mayer-Hasselwander}}
  \bibnamefont{and} \bibinfo{author}{\bibnamefont{et~al.}},
  \bibinfo{journal}{A\& A} \textbf{\bibinfo{volume}{335}}, \bibinfo{pages}{161}
  (\bibinfo{year}{1998}).

\bibitem[{\citenamefont{Biller and et~al.}(1998)}]{Biller98}
\bibinfo{author}{\bibfnamefont{S.~D.} \bibnamefont{Biller}} \bibnamefont{and}
  \bibinfo{author}{\bibnamefont{et~al.}}, \bibinfo{journal}{Phys. Rev. Lett.}
  \textbf{\bibinfo{volume}{80}}, \bibinfo{pages}{2992} (\bibinfo{year}{1998}).

\bibitem[{\citenamefont{Vassiliev}(2000)}]{Vassiliev00}
\bibinfo{author}{\bibfnamefont{V.~V.} \bibnamefont{Vassiliev}},
  \bibinfo{journal}{Astroparticle Physics} \textbf{\bibinfo{volume}{12}},
  \bibinfo{pages}{217} (\bibinfo{year}{2000}).

\bibitem[{\citenamefont{Primack et~al.}(1999)\citenamefont{Primack, Bullock,
  Somerville, and MacMinn}}]{Primack99}
\bibinfo{author}{\bibfnamefont{J.~R.} \bibnamefont{Primack}},
  \bibinfo{author}{\bibfnamefont{J.~S.} \bibnamefont{Bullock}},
  \bibinfo{author}{\bibfnamefont{R.~S.} \bibnamefont{Somerville}},
  \bibnamefont{and} \bibinfo{author}{\bibfnamefont{D.}~\bibnamefont{MacMinn}},
  \bibinfo{journal}{Astroparticle Physics} \textbf{\bibinfo{volume}{11}},
  \bibinfo{pages}{93} (\bibinfo{year}{1999}).

\bibitem[{\citenamefont{Hawking}(1975)}]{hawk74}
\bibinfo{author}{\bibfnamefont{S.~W.} \bibnamefont{Hawking}},
  \bibinfo{journal}{Nature} \textbf{\bibinfo{volume}{248}}, \bibinfo{pages}{30}
  (\bibinfo{year}{1975}).

\bibitem[{\citenamefont{MacGibbon and Webber}(1990)}]{macgibbon}
\bibinfo{author}{\bibfnamefont{J.}~\bibnamefont{MacGibbon}} \bibnamefont{and}
  \bibinfo{author}{\bibfnamefont{B.~R.} \bibnamefont{Webber}},
  \bibinfo{journal}{Phys Rev. D} \textbf{\bibinfo{volume}{41}},
  \bibinfo{pages}{3052} (\bibinfo{year}{1990}).

\bibitem[{\citenamefont{Heckler}(1997{\natexlab{a}})}]{heckler1}
\bibinfo{author}{\bibfnamefont{A.~F.} \bibnamefont{Heckler}},
  \bibinfo{journal}{Phys. Rev. D} \textbf{\bibinfo{volume}{55}},
  \bibinfo{pages}{480} (\bibinfo{year}{1997}{\natexlab{a}}).

\bibitem[{\citenamefont{Heckler}(1997{\natexlab{b}})}]{heckler2}
\bibinfo{author}{\bibfnamefont{A.~F.} \bibnamefont{Heckler}},
  \bibinfo{journal}{Phys. Rev. Lett.} \textbf{\bibinfo{volume}{78}},
  \bibinfo{pages}{3430} (\bibinfo{year}{1997}{\natexlab{b}}).

\bibitem[{\citenamefont{Cline et~al.}(1999)\citenamefont{Cline, Mostoslavsky,
  and Servant}}]{cline}
\bibinfo{author}{\bibfnamefont{J.~M.} \bibnamefont{Cline}},
  \bibinfo{author}{\bibfnamefont{M.}~\bibnamefont{Mostoslavsky}},
  \bibnamefont{and} \bibinfo{author}{\bibfnamefont{G.}~\bibnamefont{Servant}},
  \bibinfo{journal}{Phys Rev D} \textbf{\bibinfo{volume}{59 no. 6}},
  \bibinfo{pages}{630091/1} (\bibinfo{year}{1999}).

\bibitem[{\citenamefont{Daghigh and Kapusta}(2001)}]{kapusta}
\bibinfo{author}{\bibfnamefont{R.~G.} \bibnamefont{Daghigh}} \bibnamefont{and}
  \bibinfo{author}{\bibfnamefont{J.~I.} \bibnamefont{Kapusta}}
  (\bibinfo{year}{2001}), \eprint{gr-qc/0109090}.

\bibitem[{\citenamefont{Amelino-Camelia and et~al.}(1998)}]{amelino98}
\bibinfo{author}{\bibfnamefont{G.}~\bibnamefont{Amelino-Camelia}}
  \bibnamefont{and} \bibinfo{author}{\bibnamefont{et~al.}},
  \bibinfo{journal}{Nature} \textbf{\bibinfo{volume}{393}},
  \bibinfo{pages}{763} (\bibinfo{year}{1998}).

\bibitem[{\citenamefont{Biller and et~al.}(1999)}]{biller}
\bibinfo{author}{\bibfnamefont{S.~D.} \bibnamefont{Biller}} \bibnamefont{and}
  \bibinfo{author}{\bibnamefont{et~al.}}, \bibinfo{journal}{Phys. Rev. Lett.}
  \textbf{\bibinfo{volume}{83 no. 11}}, \bibinfo{pages}{2108}
  (\bibinfo{year}{1999}).

\bibitem[{\citenamefont{Ellis and et~al.}(2000)}]{ellis}
\bibinfo{author}{\bibfnamefont{J.}~\bibnamefont{Ellis}} \bibnamefont{and}
  \bibinfo{author}{\bibnamefont{et~al.}}, \bibinfo{journal}{Astrophys. J.}
  \textbf{\bibinfo{volume}{535}}, \bibinfo{pages}{139} (\bibinfo{year}{2000}).

\bibitem[{\citenamefont{Gaidos and et~al.}(1996)}]{Gaidos96}
\bibinfo{author}{\bibfnamefont{J.~A.} \bibnamefont{Gaidos}} \bibnamefont{and}
  \bibinfo{author}{\bibnamefont{et~al.}}, \bibinfo{journal}{Nature}
  \textbf{\bibinfo{volume}{383}}, \bibinfo{pages}{319} (\bibinfo{year}{1996}).

\bibitem[{\citenamefont{Cawley and et~al.}(1990)}]{cawley90}
\bibinfo{author}{\bibfnamefont{M.~F.} \bibnamefont{Cawley}} \bibnamefont{and}
  \bibinfo{author}{\bibnamefont{et~al.}}, \bibinfo{journal}{Exp. Astron.}
  \textbf{\bibinfo{volume}{1}}, \bibinfo{pages}{173} (\bibinfo{year}{1990}).

\bibitem[{\citenamefont{Barrau and et~al.}(1998)}]{barrau98}
\bibinfo{author}{\bibfnamefont{A.}~\bibnamefont{Barrau}} \bibnamefont{and}
  \bibinfo{author}{\bibnamefont{et~al.}}, \bibinfo{journal}{Nucl. Instrum.
  Methods A} \textbf{\bibinfo{volume}{416}}, \bibinfo{pages}{278}
  (\bibinfo{year}{1998}).

\bibitem[{\citenamefont{Hara and et~al.}(1993)}]{hara93}
\bibinfo{author}{\bibfnamefont{T.}~\bibnamefont{Hara}} \bibnamefont{and}
  \bibinfo{author}{\bibnamefont{et~al.}}, \bibinfo{journal}{Nucl. Instrum.
  Methods A} \textbf{\bibinfo{volume}{332}}, \bibinfo{pages}{300}
  (\bibinfo{year}{1993}).

\bibitem[{\citenamefont{Daum and et~al.}(1997)}]{daum97}
\bibinfo{author}{\bibfnamefont{A.}~\bibnamefont{Daum}} \bibnamefont{and}
  \bibinfo{author}{\bibnamefont{et~al.}}, \bibinfo{journal}{Astropart. Phys.}
  \textbf{\bibinfo{volume}{8}}, \bibinfo{pages}{1} (\bibinfo{year}{1997}).

\bibitem[{\citenamefont{Sinnis}(2001)}]{sinnis01}
\bibinfo{author}{\bibfnamefont{C.}~\bibnamefont{Sinnis}}, in
  \emph{\bibinfo{booktitle}{in {\it Proc. of the 27th ICRC}}}
  (\bibinfo{address}{Hamburg, Germany}, \bibinfo{year}{2001}), p.
  \bibinfo{pages}{(in press)}.

\bibitem[{\citenamefont{Bussino and Mari}(2001)}]{bussino}
\bibinfo{author}{\bibfnamefont{S.}~\bibnamefont{Bussino}} \bibnamefont{and}
  \bibinfo{author}{\bibfnamefont{S.~M.} \bibnamefont{Mari}},
  \bibinfo{journal}{Astropart. Phys.} \textbf{\bibinfo{volume}{15}},
  \bibinfo{pages}{65} (\bibinfo{year}{2001}).

\bibitem[{\citenamefont{Piazzoli}(2001)}]{Piazzoli}
\bibinfo{author}{\bibfnamefont{B.~D.} \bibnamefont{Piazzoli}},
  \bibinfo{journal}{Private Communication}  (\bibinfo{year}{2001}).

\end{thebibliography}

\end{document}